\documentclass[structabstract]{aa}
\usepackage{natbib}
\usepackage{amsmath}
\usepackage{graphicx}
\usepackage{txfonts}

\newcommand{\degree}{^\circ}
\newcommand{\ud}{{\mathrm d}}
\usepackage{xargs}                      
\newcommand{\toprule}{ \hline \hline \\ [-1ex]}
\newcommand{\midrule}{\\ [-1ex] \hline \\ }
\newcommand{\bottomrule}{\\ [-1ex] \hline \\ }
\begin{document}

\title{
  A transmission hologram for slitless spectrophotometry on a convergent telescope beam.
  }
\subtitle{
  II. Optimisation and characterization.
  }
\titlerunning{A transmission hologram for slitless spectrophotometry on a convergent telescope beam.2.}

\author{
S.~Dagoret-Campagne\inst{1} \and
M.~Moniez\inst{1} \and
J.~Neveu\inst{1} \and
A.~Blot\inst{1} \and
P.~Antilogus\inst{2} \and
C.~Juramy\inst{2} \and
L.~Le Guillou\inst{2} \and
Ph.~Repain\inst{2} \and
E.~Sepulveda\inst{2} \and
C. Michel\inst{3} \and
F. Colas\inst{4}
}
\institute{
Laboratoire de Physique des deux infinis Ir\`ene Joliot-Curie, Universit\'e Paris-Saclay, CNRS/IN2P3, IJCLab, 91405 Orsay, France
Sorbonne Universit\'e, CNRS/IN2P3 \and
Laboratoire de Physique Nucl\'eaire et de Hautes \'Energies (LPNHE),
75005 Paris, France \and
IP2I - Plateforme LMA, Bât. VIRGO
7, av Pierre de Coubertin - Villeurbanne 69100, France \and
Institut de m\'ecanique c\'eleste et de calcul des \'ephém\'erides,
Observatoire de Paris, 77 avenue Denfert-Rochereau, F-75014, Paris, France
}

\date{Received 29/09/2025, accepted 30/09/2026}
\abstract
{
This article details the optimization and the characterization of the hologram described in
a companion paper published in 2021, which showed the superiority of a holographic optical element over a periodic grating as a disperser installed in the path of a converging beam on an on-axis detector (unbent spectrograph) for slitless spectroscopy.
This hologram now equips the auxiliary telescope (AuxTel) at the Rubin-LSST observatory, a refurbished telescope designed to serve as both an imager and a spectrophotometer for measuring the transparency of the atmosphere as a function of wavelength.
}
{
We describe in detail the steps and results of optimizing the final optical holographic element installed on the spectrograph, with an emphasis on the equipment and modeling tools used.
The goal was to produce a hologram with the best diffraction performance across the entire spectrum covered by the filters of the Rubin-LSST observatory's main telescope.
}
{
We utilized our technical resources — an optical bench and telescopes for sky measurements — as well as specialized modeling tools, which enabled us to determine the optimal production parameters for the AuxTel hologram after four prototyping phases.
We also conducted on-sky checks and measurements using various telescopes.
}
{
Thanks to this optimization, we have succeeded in achieving a maximum first-order diffraction transmission of $32\%$ with our thin hologram, a value very close to the theoretical maximum possible.
This hologram has been in place on AuxTel's spectrograph since February 2021
and has been performing satisfactorily ever since, in conjunction with analysis software adapted to slitless spectroscopy.
}
{}

\maketitle
\nolinenumbers
\section{Introduction}

The Vera Rubin Observatory operates the Simonyi Survey Telescope (SST), a 8.4 meter diameter telescope, equipped with a 3.2 Gpixels
camera, for the Legacy Survey of Space and Time (LSST), a 10 year south sky survey through 6 wide-band filters ugriZy
\citep{LSSTScienceBook_2009}. The objective of the standard calibration
procedure is to reach a photometric precision of $0.5\%$, but the most complete calibration scheme should allow us to improve this down to $0.1\%$ \citep{Stubbs_2007}, by taking real-time atmospheric transmission into account.
Knowing this atmospheric transmission will enable us to deduce in real time the effective pass-band crossed by the light between the top of the atmosphere and the detector output.
The main objective is to bring LSST's broadband photometry as close as possible, star by star and exposure by exposure, to standard atmospheric conditions, in order to achieve photometric repeatability in the milli-magnitude range.
To reach this precision for each LSST color band, best quality visible spectra from near UV to near IR are a major asset.
The Vera Rubin Observatory auxiliary telescope (AuxTel),
an old refurbished imager telescope (diameter 1.2~m, $f/18$, scale factor at focal plane $105~\mu$m/arcsec), installed on the "calibration hill" near the main telescope, has been adapted for this purpose.
It is equipped with a slitless spectrograph, specifically dedicated to the regular measurement of local atmospheric transmission through spectrophotometric monitoring of standard stars between $345\,$nm and $1050\,$nm.
slitless spectroscopy enables for approximate, and therefore rapid, pointing, a decisive advantage for measuring atmospheric transmission in real time in a direction close to that of the main telescope.
A key feature of this spectrograph was that it had to be capable of being mounted on the same beam path as an imager, with the ability to switch quickly from the imager configuration to the spectrograph configuration. Consequently, this spectrograph cannot be angled—which produces optical path length variations with wavelength—and its disperser must be inserted into a converging (rather than parallel) beam. The hologram-based technique we propose has thus led us to a solution that can be generalized to any imager that the user might wish to easily convert into a spectrophotometer.

This article is a follow-up to the paper \citet{holospec1} (hereafter called paper 1), which demonstrated the superiority of thin hologram focusing over standard periodic gratings for transforming AuxTel imager into an unbent slitless low-resolution spectrograph by simply interposing the hologram in the telescope's converging beam.
The relevant elements of this hologram are, on the one hand, its optical function to ensure the best focusing on the whole spectrum (studied in detail in paper 1), on the other hand, its large first-order diffraction transmission.

\begin{figure} 
\begin{center}
\includegraphics[width=8cm]{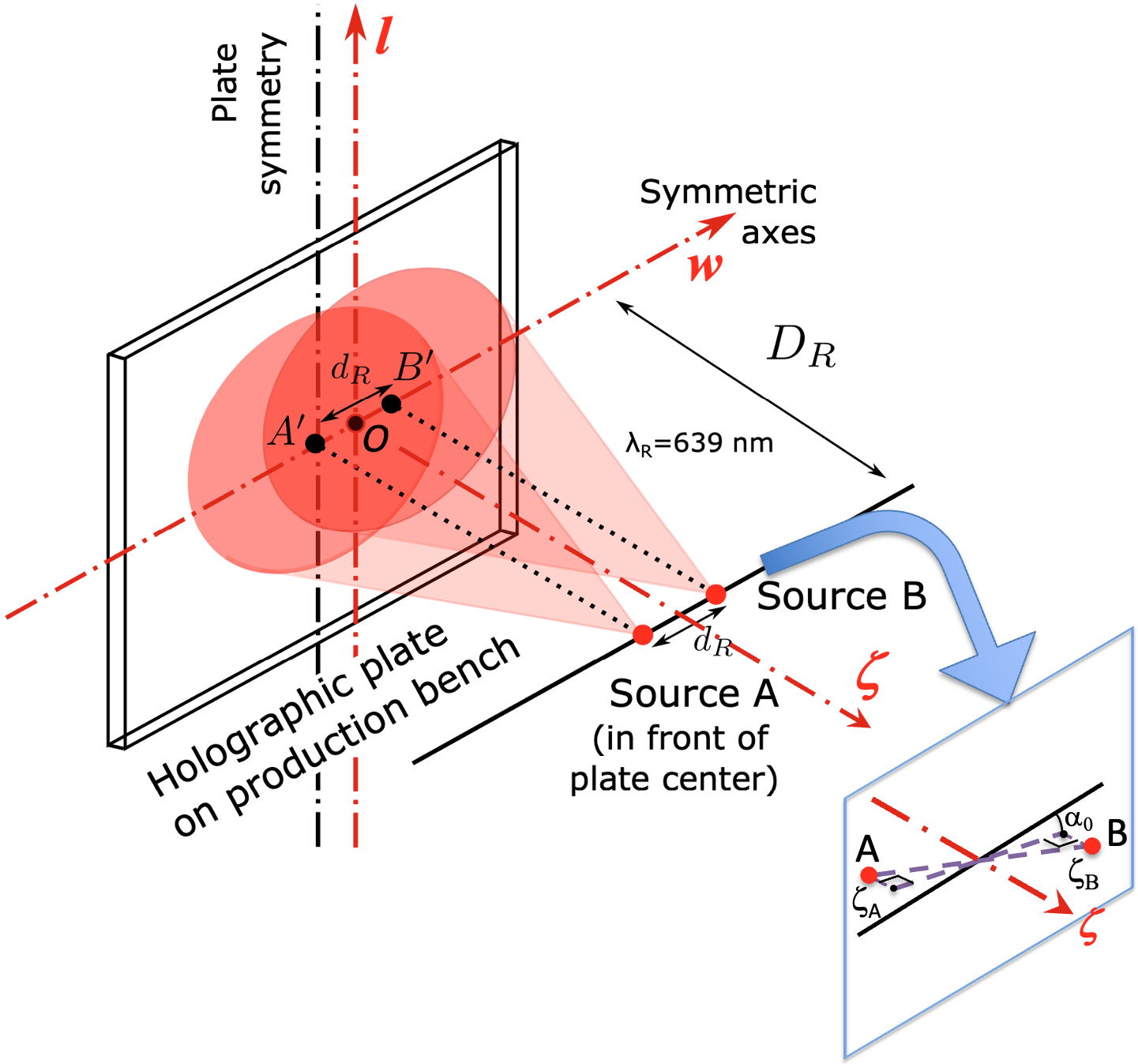}\\
\end{center}
\caption[] 
{
Recording of our specific hologram.
The intensity interference pattern of two illuminating point sources A (reference wave) and B (image wave), produced by a laser split beam ($\lambda_R=639\,$nm), is recorded on the holographic sensitive plate emulsion.
The insert at the right is an enlarged view showing the parameters for the positions of sources A and B as used in section \ref{Sect:dispersion}.

$(O,w,l,\zeta)$ is the coordinate referential attached to the hologram. $(w,l)$ axis
are the symmetric axis of the interference pattern, and the black dashed lines are centered on the incident telescope beam.
}
\label{prod-holo}
\end{figure}

Conventional periodic gratings used in on-axis spectrographs (in which the detector plane is kept perpendicular to the telescope axis) introduce wavelength-dependent optical path variations and image distortions when illuminated by a converging beam, leading to degraded focusing. Holographic gratings mitigate these effects through their non-strict periodicity, which provides an additional focusing capability. These imaging properties have been validated by preliminary tests at the CTIO 0.9-m telescope (paper 1).
Figure \ref{prod-holo} shows the principle of realization of the hologram.
When illuminated by a converging wave (at point A) matching the recording geometry, the hologram produces a point-like first-order diffracted image on the line $AB$, whose position follows a dispersion relation governed by the local effective line density.

In this work, assuming near-normal incidence, we focus on the transmission properties and on the optimization of the hologram manufacturing process.

Since our first publication we have optimized the hologram from the initial prototypes -- adapted to the CTIO telescope -- to the final object that is now installed on the AuxTel, following the study of three more sets of prototypes.

The aim of the present paper is to describe this optimization of the hologram production process (type of emulsion, exposure time, relative intensity of interfering beams, etc.) in order to maximize the first-order diffracted intensity of light from a point source focused at the telescope's focal plane.
We have chosen to describe our approach and procedures in sufficient detail because we believe that they can be applied to enable any imaging telescope to be transformed into a slitless spectrophotometer simply by inserting a thin hologram into the converging beam path at the focal point, without any other mechanical (bending mechanism) or optical modifications.

The paper is organized as follows:
In Sect. \ref{Sect:optproc} we discuss the room for maneuver with the hologram producing techniques, focusing on the procedures that are controlled by our hologram maker,
on the parameters that are relevant to our optimization, and to the specific geometry of the AuxTel. 
In Sect. \ref{Sect:bench}, we describe the optical test-bench we have used, and specifically the development of a telescope optical beam simulator. This bench has been used to measure the imaging and diffraction transmission characteristics of all the holograms in a controlled situation.
A basic model of diffraction by a phase-modulated grating was used to guide the production of the series of prototypes described in Section \ref{Sect:simulation}.
A more sophisticated model, allowing us to simulate both the hologram production and the image restitution has been developed to understand the parameters responsible for the transmission efficiencies of the final hologram (Sect. \ref{Sect:model}).
The characteristics and performances of the final holograms are described in Sect. \ref{Sect:performance}, and compared with the model.
In Sect. \ref{Sect:pic}, we describe some early measurements made on the sky with the one-meter telescope of the Pic du Midi observatory, which enabled us to carry out “life-size” tests in a real situation, thanks to the similarity of the geometric configuration with that of the AuxTel.
The commissioning of AuxTel, robustness tests, and late improvements to the spectrograph are discussed in the remaining part of Sect. \ref{Sect:discussion}, before concluding (Sect. \ref{Sect:conclusion}).

\section{Adapting and optimizing the process}
\label{Sect:optproc}
Our aim was to optimize the hologram manufacturing process so as to maximize transmission in the first diffraction order.
We decided to concentrate on phase holograms and to discard amplitude holograms, on the basis of the tests done for paper 1 on the 1m CTIO telescope which showed that the former had a much better diffraction transmission.

\begin{figure} 
\begin{center}
\includegraphics[width=7.cm]{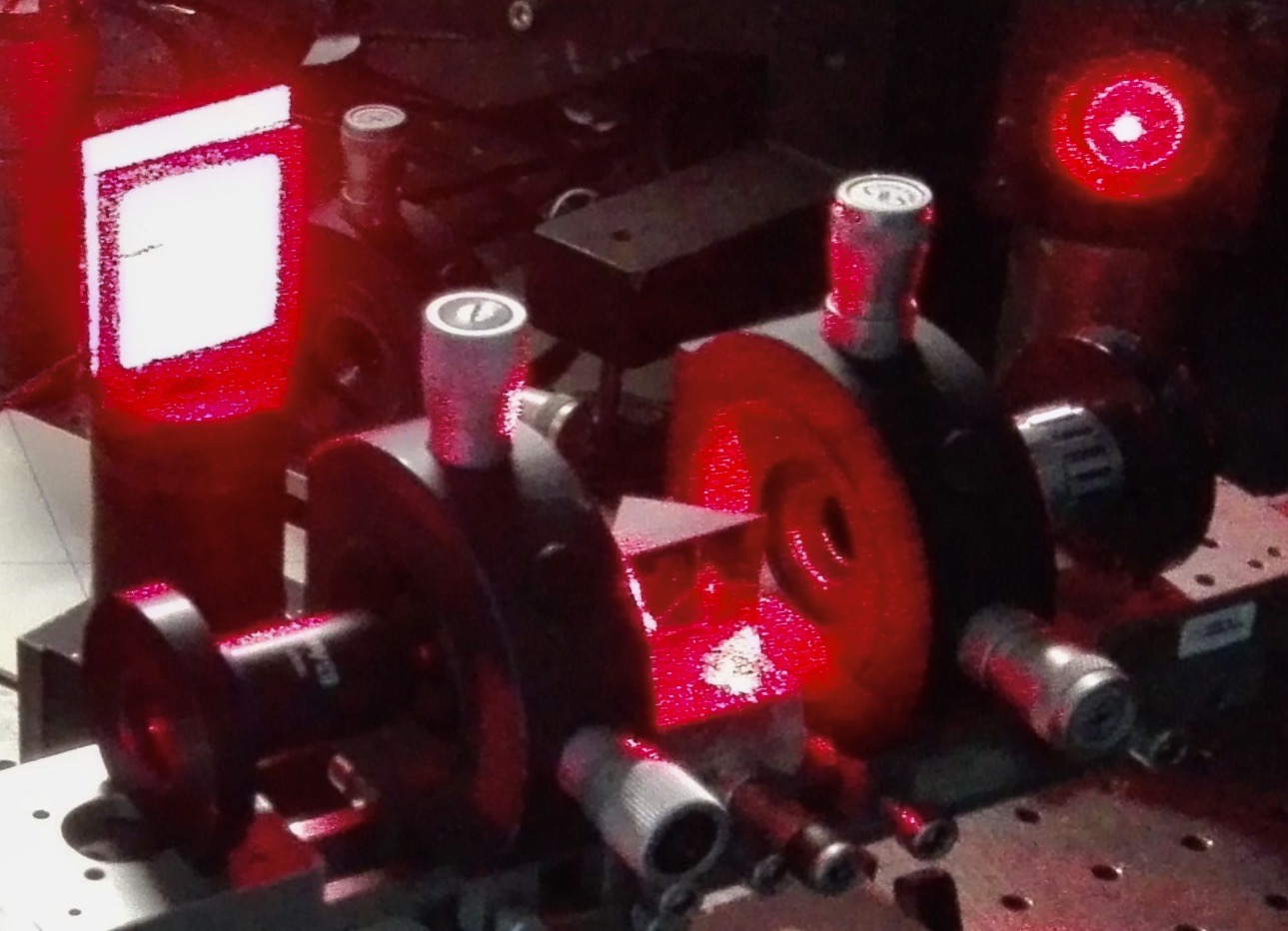}
\end{center}
\caption[] 
{
Detail of the assembled production optical bench showing the microscope objectives, the $30\mu$m filtering holes, the prism-mirror and the illuminated plate.
}
\label{bench-prod}
\end{figure}

First and foremost, the holographic production bench was modified and upgraded from the time of the first prototypes made for the CTIO telescope.
In the geometric configuration of the AuxTel, the distance between $A$ and $B$ sources is $d_\text{R}=20$mm, allowing us to insert microscope objectives in front of the $30\mu$m filtering holes creating the A and B sources (see Fig. \ref{bench-prod}).
As a consequence, parasitic light was better suppressed, avoiding beating effects (Moir\'e figures) in the recorded fringe system, that appeared on the prototypes adapted to the CTIO.
This geometrical configuration produced a figure of diffraction with a line density
$N_\text{eff}\sim 150$ lines/mm at the optical center $A'$ (inter-fringe spacing in mm $a=1/N_\text{eff}$).

After exposure to the interference pattern of the sources, the emulsions of the prototypes and final holograms
where chemically processed (wet processing) with a sequence of baths : developing, stopping, fixing and bleaching baths. The bleaching bath made the Holograms transparent (from UV to IR), and allowed us to produce phase-modulated holograms thanks to the local refraction index modulations, where the highly illuminated zones have larger refraction index after processing.

The optimization strategy, detailed in Sect. \ref{Sect:simulation}, was driven by optical test-bench measurements, which enabled a progressive refinement of the fabrication parameters (e.g. exposure time, grain size, and chemical bath composition) over three successive generations of prototypes. This iterative approach allowed us to approach the theoretical maximum first-order transmission of $33.9\%$ expected for thin\footnote{Our holograms are thin because the thickness of the recording medium ($e\sim 5\mu\text{m}$) is such that the fringe shift within this thickness is much smaller than the recorded inter-fringe spacing $a$. This shift is quantified by the variation of the phase difference between the interfering beams within the thickness of the emulsion, with the $Q-parameter$ defined as (\cite{Kogelnik_1969,Bjelkhagen_1995}, p61)
$Q=2\pi.e.\lambda_\text{R}/(n_\text{e} a^2)$,
where $\lambda_\text{R}$ is the recording laser wavelength and $n_\text{e}$ is the refraction index of the emulsion.
In our case, $\text{Q}\sim 0.30$, smaller than $1$, characterizing a thin hologram \citep{Bjelkhagen_1995}.}
linear phase holograms, assuming a quasi-linear variation of the phase delay (or optical path difference) with illumination \citep{Dammann_1970}.

\section{Laboratory measurements}
\label{Sect:bench}
To study prototypes at different stages of optimization, we took advantage of an optical test bench,
which we equipped with a telescope beam simulator.

\subsection{The optical test-bench}
\begin{figure*}[ht!]
\begin{center}
\includegraphics[width=0.9\linewidth]{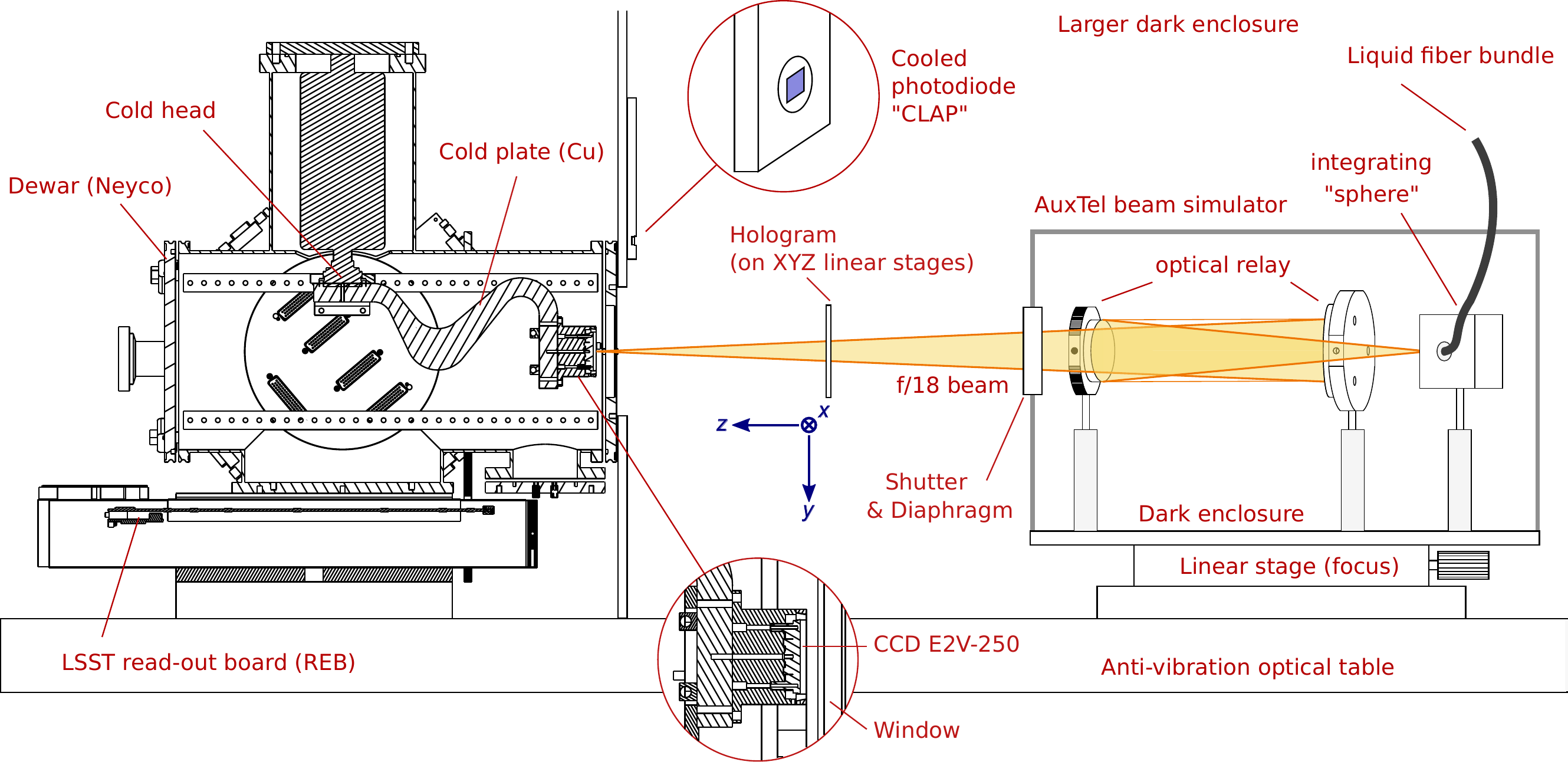}
\caption{Optical test-bench (side view). The Quartz Tungsten Halogen lamp and the monochromator sit above the dark enclosure and are not drawn. The light from the monochromator exit slit is conveyed through a liquid light guide (Newport 77639 Liquid Light Guide, 2~m long, 8~mm diameter) that feeds the small (cubic) integrating sphere. Exposure time is controlled through a mechanical shutter attached at the exit of the optical beam simulator (see below).}
\label{fig:test-bench}
\end{center}
\end{figure*}

This test-bench (see Fig. \ref{fig:test-bench}) has been initially designed to qualify the sensors
for the CCD camera of the Vera Rubin LSST telescope and to optimize
their readout procedures (see \citealt{JuramySPIE2014} and
\citealt{OConnorSPIE2016}). For the hologram tests, we performed the
measurements with a
STA/ITL fully depleted CCD developed for the
LSST project \citep{Lesser_2017}.
This sensor, made of high-resistivity silicon $100\mu$m thick, has a $10\mu$m pixel side, and 4096x4004 pixels in total. It is divided into 16 segments, each of which is
512x2002 pixels in size and has its own readout channel. The CCD is kept inside a Neyco Dewar at pressure below $7.10^{-7}$mb and
temperature controlled at $-100.00\pm0.01^{\circ}\text{C}$.
Following the vendor recommendations, we operate the sensor in full depletion mode. The CCD is read out using the LSST electronic chain \citep{JuramySPIE2014,OConnorSPIE2016}, which runs at room temperature on our test stand, in a dedicated class~10000 / ISO-7 clean room. The clean room temperature is regulated ($\pm0.2^{\circ}$C) to minimize temperature effects on the readout electronics.

\subsection{The AuxTel optical beam simulator}
\label{Sect:beam-simulator}
\begin{figure}
\begin{center}
\includegraphics[width=0.9\linewidth]{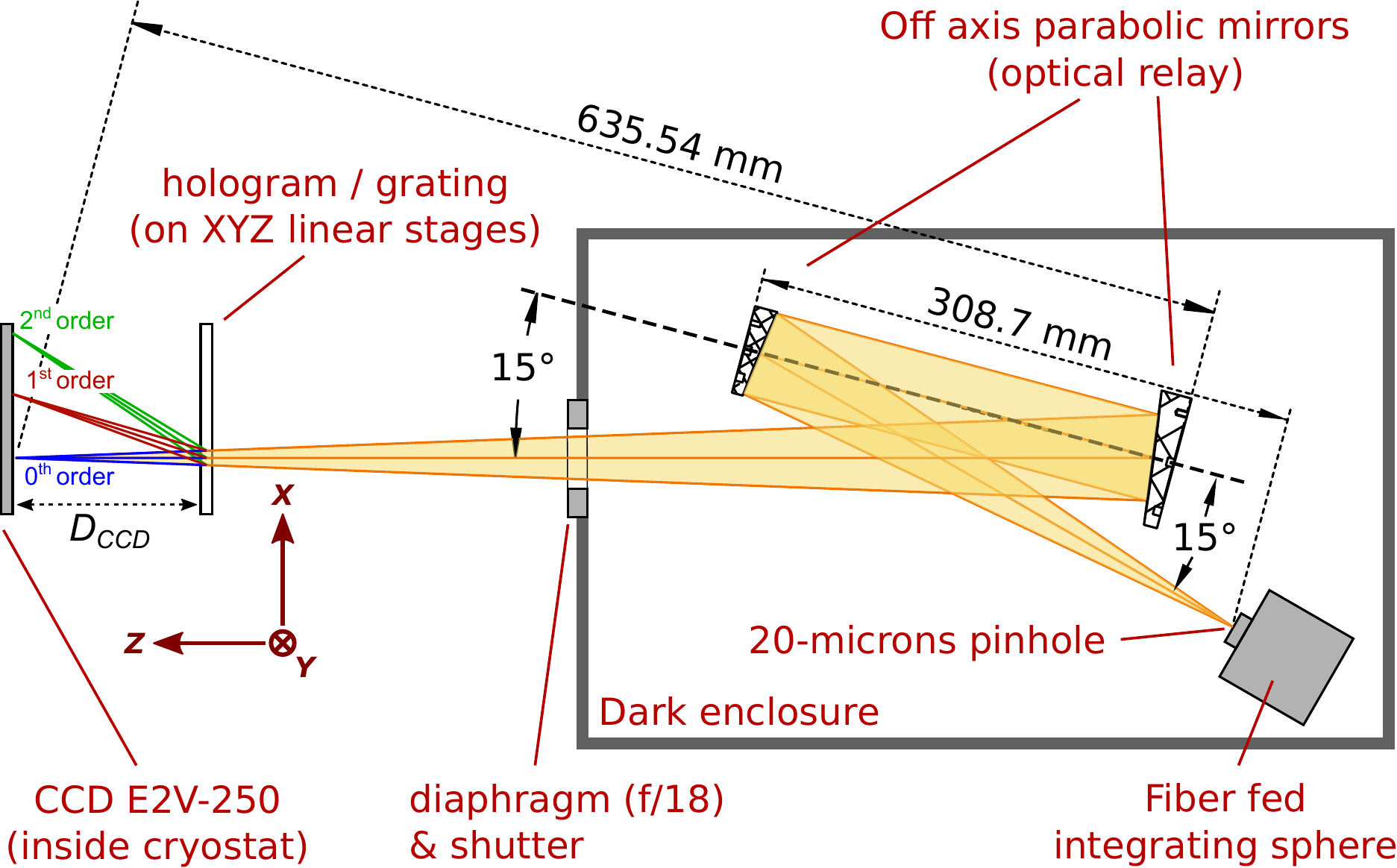}
\caption{The AuxTel optical beam simulator (see text). The light
  emerges from the integrating sphere through a $20\,\mu\text{m}$
  diameter pinhole, hits successively two off-axis parabolic
  mirrors (optical relay), then converges on the CCD to form
  point-like image with a $f/D=18$ converging beam.}
  \label{fig:beam-simulator}
\end{center}
\end{figure}

To measure the hologram characteristics with an optical configuration identical to the AuxTel, we have designed and built a specific optical system that produces a convergent uniform beam with the same aperture as AuxTel ($f/18$, see fig.~\ref{fig:beam-simulator}). From the
luminous source to the image position, the light issued from a fiber or from a Quartz Tungsten Halogen lamp passes successively though:
\begin{itemize}
    \item A small fiber-fed integrating sphere that homogenizes the directions of the rays entering the exit pin-hole. This provides an isotropic distribution of the light at the entrance of the pin-hole that defines the point-source.
    \item The pin-hole is a disk of $20\,\mu$m diameter, which is our best compromise between the output intensity and the size of the source and its image. 
    \item The first off-axis parabolic mirror (focal 310.08mm, diameter 50.8mm, 15$^\text{o}$ off-axis, model~35542 from Edmund Optics). After reflection, the light beam is parallel.
    \item The second off-axis parabolic mirror (focal 646.0mm, diameter 76.2mm, 15$^\text{o}$ off-axis, model~35597 from Edmund Optics) produces a point image. Thanks to the absence of lenses, this combination of mirrors forms an optical relay which is a rigorously stigmatic system, and independent of the wavelength.
    \item A diaphragm to fit the $f/D=18$ aperture and a shutter.
\end{itemize}

The only geometric difference between the light beam produced by this setup and the beam expected from the AuxTel is the lack of the central obtrusion due to the secondary mirror of AuxTel. This compact setup has been assembled in a black box
for protection against external stray light, that can easily be inserted within the volume of the CCD test-bench.
Given the mirror geometry, the source-to-image magnification is 2.08, yielding a theoretical image diameter of $41.6\,\mu\text{m}$ for a $20\,\mu\text{m}$ pinhole. With an AuxTel plate scale of $105\,\mu\text{m}$ per arcsec, this simulates a stellar image of 0.4 arcsec, representative of excellent seeing conditions. In practice
the mirrors were probably too thin, since we obtained an image slightly distorted in the shape of a three-leafed clover, probably due to the three screws used to align each mirror. Fortunately, the size of the image did not exceed $50\,\mu\text{m}$ (equivalent to the size of a star with a seeing of 0.5 arcsec at the AuxTel); its shape complicates the detailed image analysis but does not affect the diffraction transmission estimates.

\subsection{Procedure for measuring hologram transmission}
Our mechanical system
allows us to move each hologram in and out of the convergent illumination beam,
and to scan their properties by varying the incident beam impact parameter on
the hologram.
With the optical bench, we studied the transmission of holograms as a function of wavelength by injecting into the integrating sphere a light from a monochromator, which delivers a triangular spectrum of width $\sim 2\,$nm
centered on the wavelength of our choice.
For each wavelength, we measured
the integrals of the diffracted fluxes in orders 0, 1 and 2 reaching the CCD (with background subtraction), successively without the hologram (measurement of direct light) and with the hologram in the beam path. The very low residual backgrounds were estimated from pixels in areas far from the beam.
The procedure was repeated several times with no significant variation in results.
We measured the transmissions of prototype holograms between 400 and $1050\,$nm in the various diffraction orders, in order to optimize the production process (Sect. \ref{Sect:prototypes}).

Transmissions measured with this test bench are plotted in Figures \ref{protos-2} and \ref{fig:simu-effic}.
The transmission ratio between order 2 and order 1, shown in red in the Fig. \ref{ratios-orders}, was measured directly on the images with the hologram in place only.
In Sect. \ref{subsec:transauxtel}, we show how we used spectra obtained on the sky with AuxTel from UV-rich stellar sources to extend these measurements to shorter wavelengths.

\section{Optimizing the diffraction transmissions}
\label{Sect:simulation}
Our hologram manufacturer ("Ultimate holography", based in Bordeaux, France) produced several series of prototypes before the final holograms, varying the emulsion type, the exposure, the intensity ratio of the interfering beams and the wavelength of the laser.
\subsection{Decision guide: a basic model}
\label{subsection:decision}
The diffraction transmission $\eta_\text{p}(\lambda)$, defined as the fraction of incident light diffracted at order $p$, was measured in order to select the most efficient production procedure for our holograms.

\begin{figure} 
\begin{center}
\includegraphics[width=6.cm]{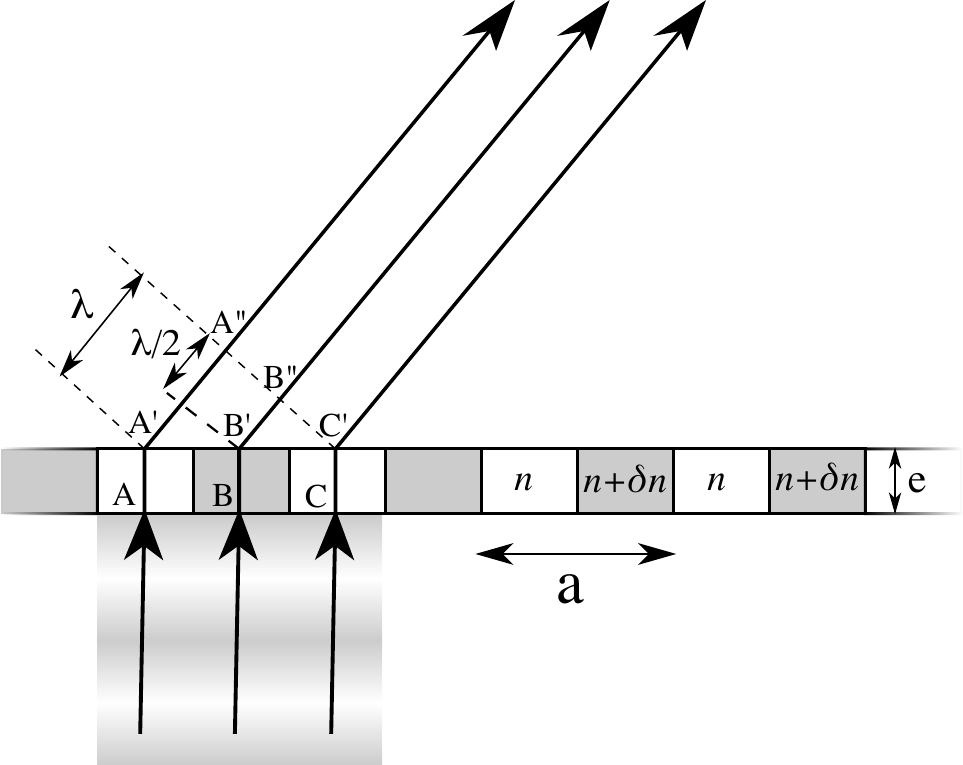}
\end{center}
\caption[] 
{
Phase grid model with periodic square-wave index variations $\delta n$, producing optical path differences $\Delta=e\delta n$. Points A, B, and C are separated by half the period (a/2).
The direction of the emerging light-rays corresponds to the first diffraction order.
Constructive interference occurs in this direction when the optical paths AA'' and BB'' match modulo $\lambda$, resulting in maximum light transmission.
}
\label{index-mod}
\end{figure}

\begin{figure} 
\begin{center}
\includegraphics[width=8.cm]{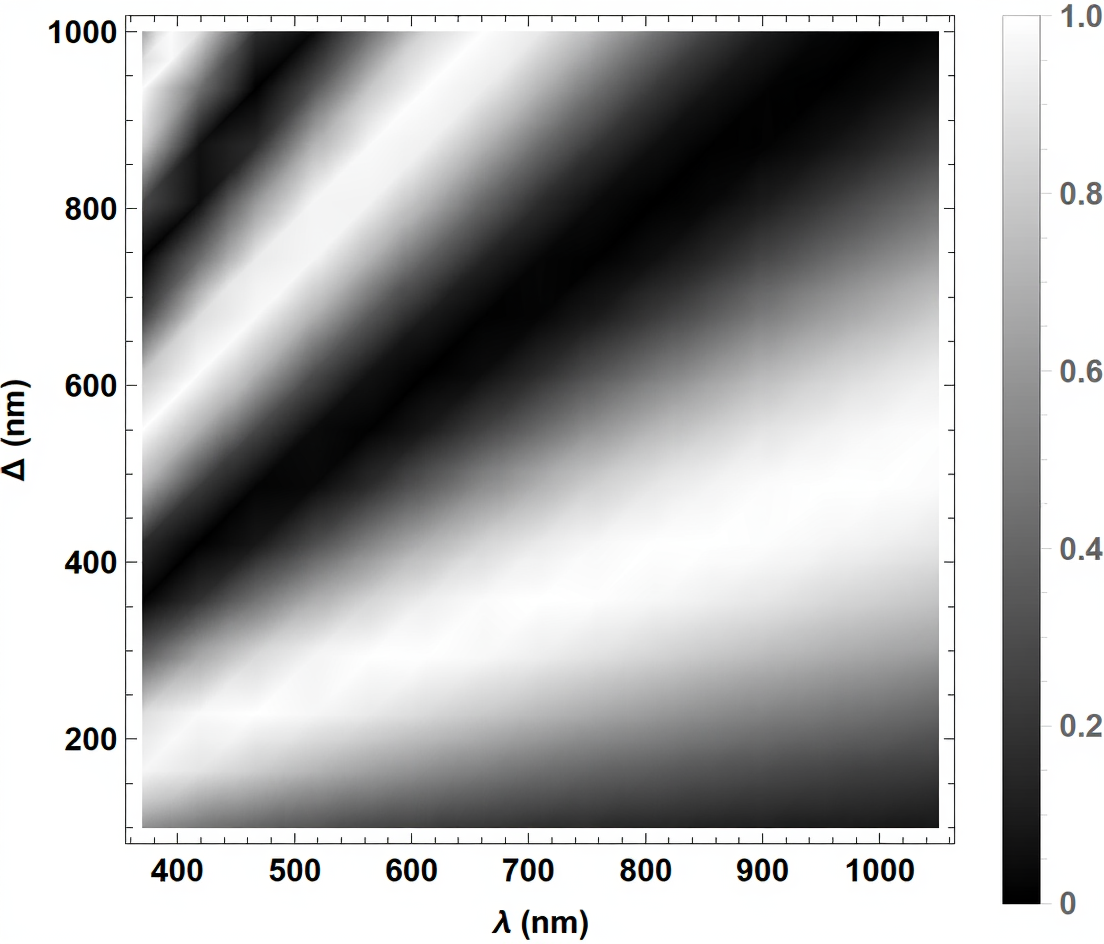}
\end{center}
\caption[] 
{
Simple model: normalized transmission of light in the first diffraction order, as a function of the wavelength $\lambda$ for optical path modulations $100\, nm<\Delta<1000\, nm$.
}
\label{efficiency}
\end{figure}

A very simple model—although it differs from our situation—helps us understand the causes of transmission variations with wavelength.
This model, illustrated on Fig. \ref{index-mod}, treating the hologram locally as a binary-phase grating with a square-wave refractive index pattern, allows us to relate the optical path modulation $\Delta$ of the hologram to the wavelengths of maxima and minima in the first-order transmission $\eta_1(\lambda)$.
Maxima occur for $\Delta=\lambda/2+k\lambda$ (where $k$ is an integer), causing the light emitted from A' and B' in the direction of the first diffraction order to interfere constructively; the zero-order light is then suppressed; the first-order minima occur when $\Delta=k\lambda$.
More generally, toward the first diffraction order, the phase difference between $A''$ and $B''$ is $\delta\phi=\pi(1-2\Delta/\lambda)$, giving an interference term $(1-\cos(2\pi\Delta/\lambda))$, which decays as $1/\lambda^2$ for $\lambda>>\Delta$.
Figure \ref{efficiency} shows the general features of the light transmission in the first order of diffraction, as a function of the wavelength, for various optical path modulations $\Delta$.
Comparison with prototype measurements indicates $200\,\text{nm}<\Delta<550\,\text{nm}$.
Considering the emulsion thickness of $5\mu$m, this corresponds to an index modulation $0.04<\delta n<0.1$, depending on the exposure, in agreement with the known maximum index modulation from the maker of the emulsion \citep{Gentet2000}.

\begin{figure} 
\begin{center}
\includegraphics[width=8.cm]{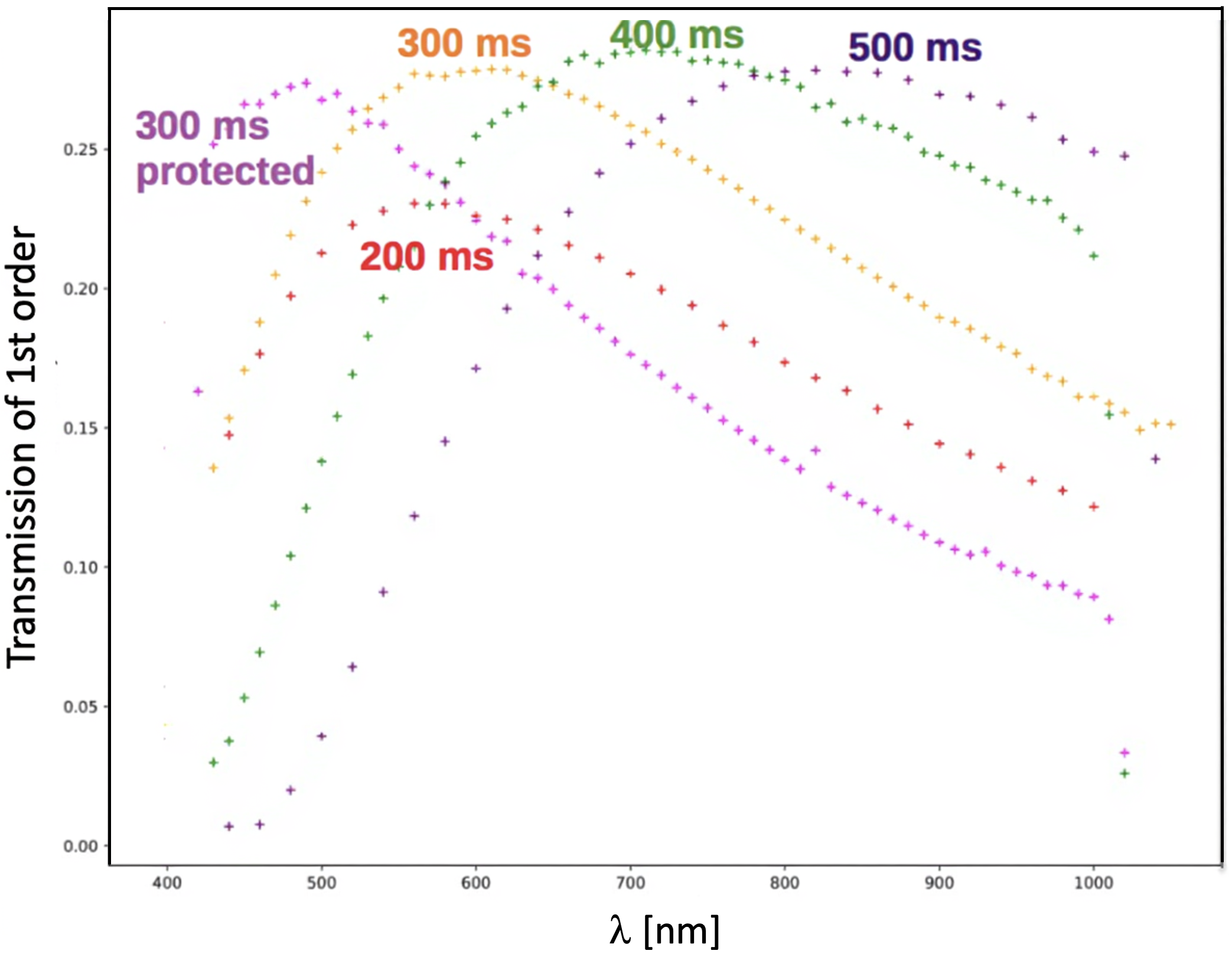}
\end{center}
\caption[] 
{
Measured transmissions $\eta_1(\lambda)$ of the light in the first diffraction order, for our second series of prototypes. The labels correspond to the various exposures; label "protected" means a sealed prototype, for which the modulation is reduced due to the suppression of the surface relief variations (see text).
The best-transmitted wavelength increases approximately linearly with exposure time, which is expected.
The poor transmission of the "200 ms" prototype is due to faulty emulsion development.
}
\label{protos-2}
\end{figure}

We have used this model to guide our decisions between subsequent prototype series, by comparing measured transmission curves.
In particular, we used the fact that the wavelength of transmission maxima increases with $\Delta$ (Fig. \ref{protos-2}).

\subsection{The three series of prototypes}
\label{Sect:prototypes}
Parameters such as the recording laser wavelength $\lambda_R$, the impact of filter holes and of the sealing, the exposure time,
the emulsion type and thickness, and the laser beam intensity ratio were adjusted to achieve the transmissions shown in Fig. \ref{fig:simu-effic}.
We optimized all these parameters to maximize first-order diffraction light over the widest wavelength range, using the series of prototypes briefly described below.

A preliminary series of tests enabled us to determine the recording laser wavelength $\lambda_\text{R}=639\,$nm, and to select the silver halide emulsions \citep{Bjelkhagen_1995}.

The first series of four prototypes was then produced on the improved optical bench with $30\,\mu$m filtering holes (Sect. \ref{Sect:optproc}).
The photosensitive plates consisted of a $5\,\mu$m thick fine-grain silver halide emulsion layer on standard BK7 glass, with
$4\,$nm grain size\footnote{the commercial denomination of this emulsion is ULTIMATE 0.4.} and $600\,\mu\text{J/cm}^2$ sensitivity \citep{Gentet2017Ultimate0T}. Each plate was positioned at $D_R=200\,$mm from the A and B sources, which were simultaneously illuminated by a split, polarized laser beam at $\lambda_R=639\,$nm.
The beam intensities were adjusted in a ratio of about 1/3 so that the resulting illumination interval fell within an approximately linear response zone of the densitometric curve of the emulsion (see the red bar on x-axis in Fig. \ref{sensitivity}).

The second series of eight prototypes allowed us to study the impact of exposure time on transmission (Fig. \ref{protos-2}). It also
included the first sealed hologram, which enabled us to quantify the effect of index-matching liquid between the hologram and the sealing glass. Sealing suppresses surface relief contributions \citep{Calixto}, reducing the phase modulation to a purely index-modulated $5\,\mu$m thick emulsion layer (with maximum relative modulation of the order of $10\%$). As a result, the wavelength of maximum first-order diffraction transmission is shifted by approximately $100\,$nm toward shorter wavelengths.

The third series of eight prototypes consisted of four pairs of sealed holograms with exposure times between 70 and $200\,$ms. For the final production, an exposure of $150\,$ms was selected as a compromise between high first-order transmission in the blue and minimal second-order transmission (see also Sect. \ref{Sect:transmission}).

\subsection{The final holograms}

The final holograms (labeled holo$\#4-$xxx) were fabricated on optical-quality BK7 glass plates (flatness $< \lambda/4$, 60/40 scratch–dig), with an anti-reflection coating on the air-facing side implemented by the Laboratoire des Matériaux Avancés (LMA) and
yielding a reflectance below $1.8\%$ per surface over 370–1050 nm. The recording exposure corresponded to a mean energy deposit of $340\,\mu\text{J/cm}^2$ ($\sim 3\,\text{mW/cm}^2$ during $\sim 150\,$ms), and parasitic reflections were suppressed using a black backing. After chemical processing, the emulsion layer was sealed with a second glass plate using the index-matching liquid; illumination during recording was performed with both plates in place to ensure stable optical path modulation.
The stability of the sensitive layer, the glue and the index-matching liquid is established to be more than 30 years, much longer than the expected Rubin-LSST operation duration.

\begin{figure} 
\begin{center}
\includegraphics[width=7.cm]{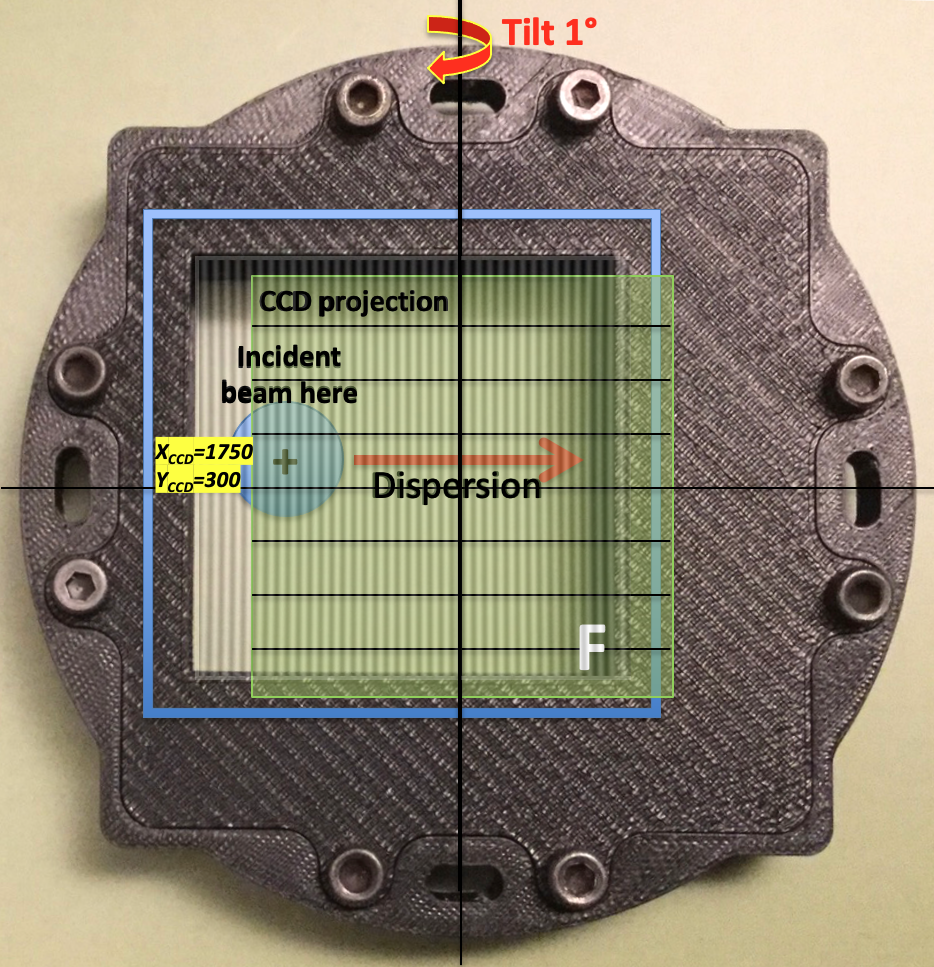}
\end{center}
\caption[] 
{
The frame of the hologram in place at AuxTel. The blue frame shows the limits of the hologram; the light green square is the projection of the CCD showing the 16 segments.
The blue disk represents the cross-section of the beam as it passes through the hologram. 
We have also superimposed the general pattern of the holographic grating. The hologram is tilted in the frame by $\sim 1 \degree$ around the vertical axis.
Note the asymmetry of the frame, which allows for several positions of the optical center relative to the CCD.
}
\label{fig:implementation}
\end{figure}

Five final hologram variants were produced. The element with the highest first-order transmission in the blue (holo$\#4$-003) was installed on AuxTel in February 2020, while a similar unit (holo$\#4$–001) is kept as a spare. The hologram is mounted in a versatile frame allowing multiple optical-axis positions on the CCD, providing operational redundancy (Fig. \ref{fig:implementation}). A $\sim 1\degree$ tilt of the hologram plane shifts internal-reflection ghosts away from the spectral region.

\section{Modeling the hologram selected for AuxTel}
\label{Sect:model}
The basic qualitative model described in section \ref{subsection:decision} served only to guide our decisions as we moved from one prototype series to the next. Once the final holograms were produced, we sought to model their transmission as accurately as possible to facilitate their use.

\subsection{Formalism}
A formal mathematical modeling of the recording of the hologram and of the image restitution is proposed in Appendix \ref{appendix_math}, using the complex 2D transmission function of the emulsion $\mathbf{H}(w,l,\lambda)$.
Other computing methods to simulate the hologram optical function exist \citep{Latychevskaia_2015}, but they are much more complex, because they are focused on the image simulation, which is not our major preoccupation in this paper.
The computational cost of these techniques would also be prohibitive, particularly for parameter optimization.
Here, we wish to establish relations between the diffracted intensity in different orders
and the parameters of the holographic production, such as
the characteristics of the exposure.
For this purpose,
we base our model (as we did for our decision guide) on the simplifying hypothesis that a holographic grating can be locally considered as a thin phase periodic grating.
We demonstrated the relevance of this approach for the imagery in paper 1, and now we use the same approximation to quantify the intensity diffracted in the different orders.
According to \cite{Song1990DiffractionEO} and \cite{CRUZ-ARREOLA2011}, the relative intensity of a plane wave of wavelength $\lambda$ diffracted at order $p$ by a grating periodic in direction $w$ (see Fig. \ref{prod-holo}), described by the (1D) periodic transfer (or transmittance) function $t(w,\lambda)$, is given by the square of the (1D) fourier coefficient $t_\text{p}(\lambda)$
\begin{equation}
    \lvert t_\text{p}(\lambda)\lvert ^2=\left\lvert\frac{1}{a}\int_0^a t(w,\lambda)\exp{[-i2\pi p w/a]}dw\right\lvert ^2,
    \label{eq:intensity}
\end{equation}
where $w$ is the coordinate along the axis perpendicular to the grating lines and $a$ is the period of the grating.
Since we use the hologram
around the optical center, at near-normal incidence,
$t(w,\lambda)=\mathbf{H}(w,0,\lambda)$,
where we approximate the local grating as a periodic one,
with line density
$N_\text{eff}\sim 150\,$lines/mm.
We assume that expression (\ref{eq:intensity}) applies to the entire section of the light beam entering the hologram, since
the hologram's inter-fringe spacing $a=1\text{mm}/N_\text{eff}$
varies by less than $0.7\%$ within the beam profile.
Even if our hologram is used with a convergent beam instead of a plane wave, considering the slow convergence and the fact that we only need the integral of the light, we can use this equation as a good approximation to study the emulsion and illumination parameters that control the profile of the recorded interference pattern.

\subsection{Modeling the hologram recording}

\subsubsection{Emulsion characteristics}

\begin{figure} 
\begin{center}
\includegraphics[width=7.cm]{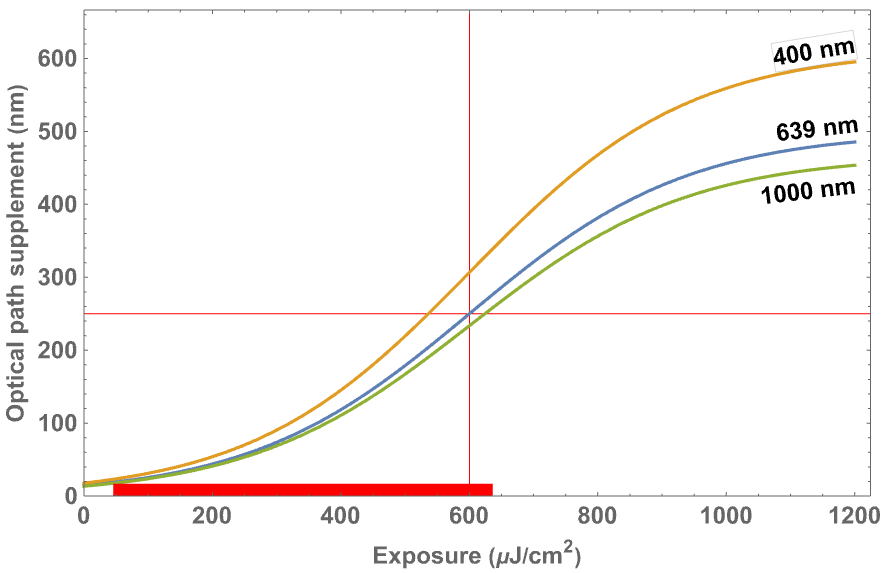}
\includegraphics[width=7.cm]{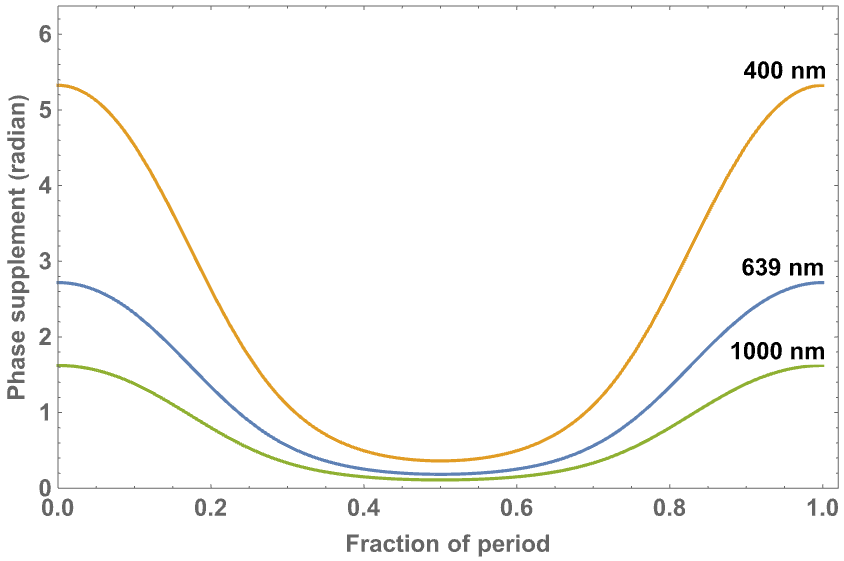}
\end{center}
\caption[]
{
(Top) the fitted characteristic curve of
$\Delta(E_\text{rec},\lambda_\text{R}=639\,nm)$
for our final hologram (blue curve).
The red horizontal bar indicates the exposure range corresponding to the interference of two beams with the characteristics shown in table \ref{tab:param} (Sect. \ref{Sect:bestfit}).
The curves labeled $\lambda=400\,nm$ and $1000\,nm$ show the effect of the emulsion exposed to $E_\text{rec}$ energy on the transmission of light at these wavelengths (see Sect. \ref{Sect:reading}).
(Bottom) the resulting phase supplement motif along the dispersion axis of the recorded periodic pattern $\Delta(E_{rec},f)$ as a function of $f$ the fractional position in the periodic motif.
Here, we show the curves corresponding to the parameters of table \ref{tab:param} that
fit the best the measured transmissions vs $\lambda$ of 0th, 1st, and 2nd orders for the final hologram.
}
\label{sensitivity}
\end{figure}

The local grating is formed by recording interference fringes from two point sources at wavelength $\lambda_R$ (Fig. \ref{prod-holo}) in a photosensitive emulsion. After development and bleaching, the resulting optical density at a position $(w,l)$, $E_\text{rec}(w,l)$ is recorded into a refractive index and thickness modulation. As mentioned in Sect. \ref{Sect:prototypes}, the sealed holograms behave as thin transmission layers with pure index modulation, within an emulsion of thickness $\sim 5\,\mu$m and maximum relative modulation of order $10\%$ (wavelength-dependent). The optical response is therefore modeled as
a complex transmittance
\begin{equation}
\mathbf{H}(w,l,\lambda)=\exp{[2i\pi\Delta(E_\text{rec}(w,l),\lambda)/\lambda]}.
\label{eq:transmission}
\end{equation}
Here, the optical path shift $\Delta(E_\text{rec}(w,l),\lambda)$ depends on both
the energy pattern recorded during manufacture and on the wavelength passing through the hologram.
At a given wavelength of use,
it is proportional to the pre-bleaching optical density of the emulsion \citep{Chang_1976},
and depends on this wavelength according to the law described in Sect. \ref{Sect:reading}.
Combined with the characteristic exposure–density response of silver-halide emulsions, this motivates a sigmoid parametrization (\citet{Bjelkhagen_1995} and \citet{Neipp_1999})
\begin{equation}
   \Delta(E_\text{rec},\lambda) = \frac{\Delta_\text{max}(\lambda)}{1+\exp{[-\gamma (E_\text{rec}-E_0)/E_0)]}},
   \label{eq:sensitometry}
\end{equation}
where
$E_0\sim 600\,\mu\text{J/cm}^2$ is the sensitivity of the Ultimate 04
emulsion \citep{Gentet2017Ultimate0T},
$\Delta_\text{max}(\lambda)$ is the maximum optical path shift at $\lambda$ when passing through the hologram (corresponding to $E_\text{rec}=\infty$),
and $\gamma$ is analogous to the photographic gamma parameter.
This model is fitted to the measured transmission of the final hologram
(Fig. \ref{sensitivity} and Sect. \ref{Sect:bestfit}).

\subsubsection{Illumination of the emulsion}
\label{subsec:illumination}
The modeling of the hologram's manufacture must now be completed by the determination of the emulsion's illumination parameters.
In our adjustment procedure, we parameterize the illumination produced by the interference of two coherent beams by the average energy deposited per unit area and the beam intensity ratio.
The determination of the parameters for the final holograms (see section \ref{Sect:bestfit}) is very close to the manufacturer's specifications.
The mathematical formalism used to describe the illumination is detailed in Annex \ref{appendix_math}1.
\subsection{Modeling the hologram reading}
\label{Sect:reading}
\begin{figure} 
\begin{center}
\includegraphics[width=7.cm]{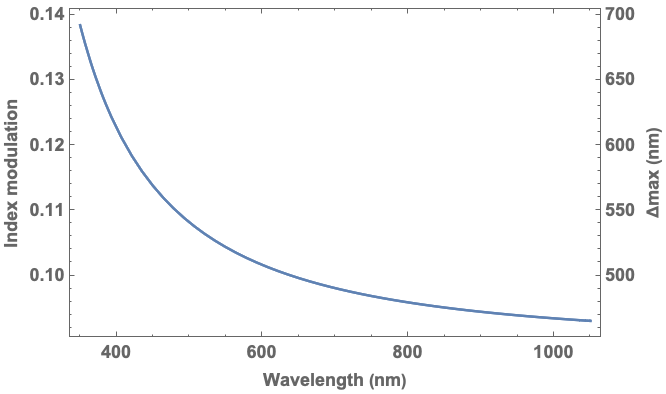}
\end{center}
\caption[] 
{
Variation in the refractive index modulation of the processed emulsion $\delta n(\lambda)$ (left scale),
and the corresponding variation in the maximum amplitude of the optical path $\Delta_{max}(\lambda)$ (right scale) when using the hologram.
}
\label{modulation-vs-lambda}
\end{figure}

As mentioned above, to model readout, we must evaluate the hologram response at any wavelength $\lambda$. This requires accounting for the wavelength dependence of the refractive index modulation $\delta n(\lambda)$.
Following \citet{Heaton_85}, we extrapolate $\delta n(\lambda)$ using the single-oscillator model of \citet{Wemple_1969}, scaled to match the properties of our emulsion \citep{Gentet2000}.
The resulting (theoretical) maximum variation of the optical path shift
$\Delta_\text{max}(\lambda)=\Delta_\text{max}(\lambda_R)\times \delta n(\lambda)/\delta n(\lambda_R)$
is shown in Fig. \ref{modulation-vs-lambda}.
For the final hologram, given the intensity ratio of the interfering beams (1/3), the effective optical path amplitude is about half this maximum.
A constant $2\%$ absorption of the emulsion is also assumed in our model.

\subsection{Degeneracies and uncertainties}
\label{Sec:degeneracies}
Our model involves several degeneracies and uncertainties. The emulsion thickness and maximum index modulation are not precisely known, but their variations can be absorbed into the fitted parameter
$\Delta_\text{max}(\lambda_\text{R})$.
Similarly, uncertainties in the emulsion sensitivity $E_0$ and in the beam intensity ratio can be incorporated into the mean exposure $\overline{E_\text{rec}}$ and the parameter $\gamma$.

Although the emulsion is highly transparent, its UV behavior is poorly constrained and likely introduces additional uncertainty. Our goal, however, is not to obtain the exact values of all parameters, but a physically consistent description that reproduces the measured diffraction efficiencies and enables the reliable fabrication of similar holographic elements.

\section{The characteristics and performances of the final holograms}
\label{Sect:performance}

Focusing and imaging characteristics were studied in details in paper 1. All the holograms tested here, and of course the one currently in place at AuxTel, have the same imaging qualities.
The spectra shown in Figs. \ref{fig:spectrogram_hg_ar}, \ref{fig:picmidi} and \ref{fig:ratiosky} illustrate in particular the quality of the focus across the entire spectral range.
In this section, we will discuss the operating range, exact dispersion, optical axis location and transmission in the different orders of the hologram in place at AuxTel.

\subsection{Fiducial domain}
Bench measurements were used to determine a conservative operating range for the hologram, which subsequently proved to be wider based on data collected from the sky
(Sect. \ref{Sect:pic}). Shifting the beam impact point in the $(w,l)$ plane (Fig. \ref{prod-holo}) shows unchanged focusing within at least
$(\Delta w,\Delta l)=\pm(5,10)\,$mm. Transmission remains constant to within $1\%$ for
$\pm(1,1)\,$mm and within $10\%$ for $\pm(4,4)\,$mm across all wavelengths.
Additional on-sky tests with varying $D_\text{CCD}$ confirm reliable operation over at least
$(\Delta w, \Delta l, \Delta D_\text{CCD})=([-5,5],[-10,10],[120,200])\,$mm around
$A'$.

\subsection{Determination of the optical center and dispersion}
\label{Sect:dispersion}
The optical center $A'$ is the optimal point where the incident light beam axis must cross the holographic grating to get optimal spectra. Its true position can be determined from the analysis of the dispersion properties as a function of the beam impact location, either from a scan of the hologram area with one light beam (on the test bench), or from multiple sources on sky (like open clusters).
An analysis of the dispersion properties of the five final holograms using the test bench also makes it possible to verify the geometry of their production retrospectively.
\begin{figure*}
    \centering
    \includegraphics[width=1.\textwidth]{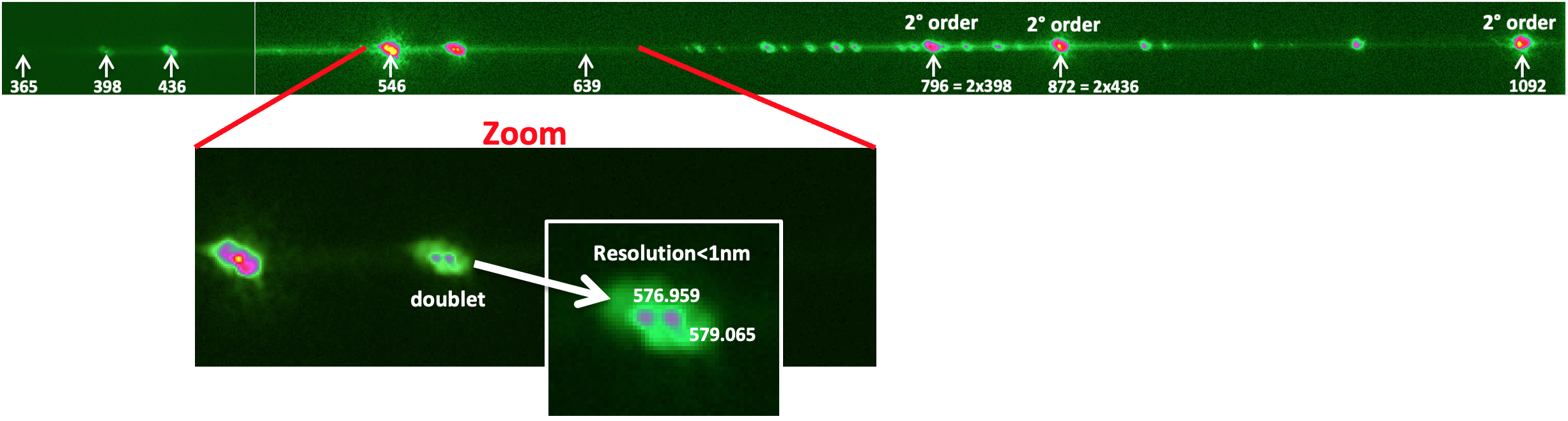}
    \caption{
    Spectrogram of $Hg-Ar$ lamp obtained with hologram Holo\#4-003 (with $N_{\mathrm{eff}} \approx 150$\,lines/mm).
    All numbers are expressed in nm. We observe the superposition of the 1st and 2nd diffraction orders.
    The lamp's fine emission lines allow us to check that the focus is unchanged across the entire spectrum.
    In particular, the inset shows that the doublet $[576.959\,\text{nm}, 579.065,\text{nm}]$ is clearly resolved.
    }
    \label{fig:spectrogram_hg_ar}
\end{figure*}

\subsubsection{Data set and spectra extraction}

With the optical test bench, we performed $(w,l)$ scans of the holograms, illuminated with a $Hg-Ar$ lamp,
adjusting $D_\text{CCD}\approx 200\,$mm (as the AuxTel configuration).
An example of spectrograms obtained with this setup is shown in Figure~\ref{fig:spectrogram_hg_ar}. The Point Spread Function (PSF) of the order 0 appeared to have a clover shape, as well as the imprints of the lamp's emission lines, due to the three screws behind the parabolic mirror (Sect. \ref{Sect:beam-simulator}). Going toward the red wavelengths, the PSF shape evolves slowly but remains clearly narrower than with a periodic grating at similar deviation from the order 0. 

To determine the optical center (A'), we first extract the centroids of the emission line images from the first diffraction order and the dispersion axis angle.
For this purpose, we built a cloverleaf-shaped PSF function from an interpolation of the order 0 image, which we reuse for the other orders.
This is sufficient to get well defined positions of the emission lines in the spectra.
Using Spectractor software \citep{Spectractor},
we determined the true (reference) distance $D_{\mathrm{CCD}} = 202.3 \pm 0.1\,$mm from the positions of spectral lines produced by a blazed Thorlabs Inc. periodic grating with exactly $300$ lines/mm installed in place of the hologram.

\subsubsection{Measurement of $N_{\mathrm{eff}}$ and dispersion angles}

\begin{figure}
    \centering
    \includegraphics[width=0.5\textwidth]{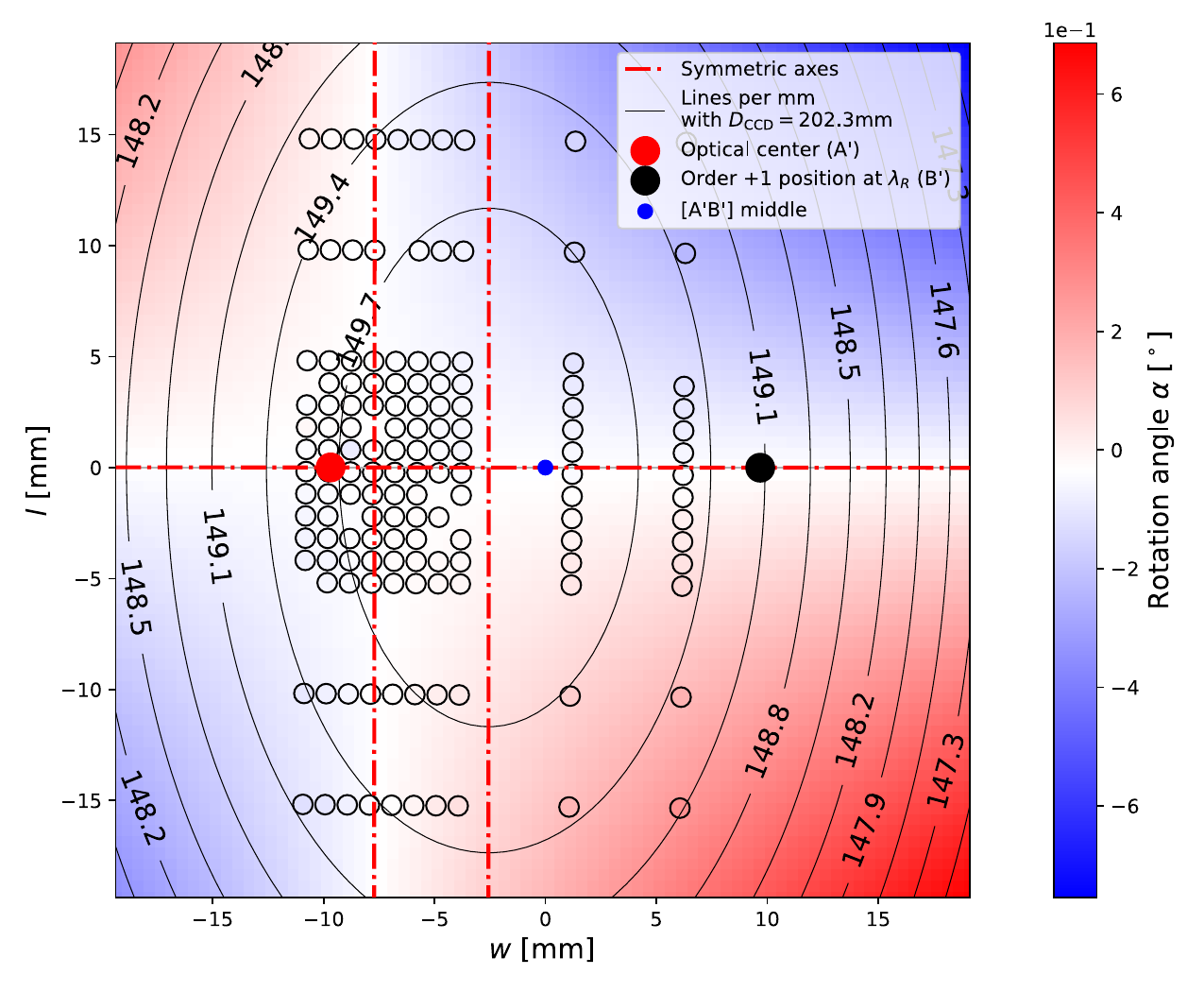}
    \caption{Best fitting model on the groove density map $N_\mathrm{eff}(w,l)$ (black contours) and on the dispersion axis angles $\alpha(w,l)$ map (colored) for the holographic grating Holo\#4-003. The circles indicate the pointings of the hologram scan, and their filling color follow the same color scale as the model.
    The vertical axis of symmetry on the left ($w\sim -8$) corresponds to the map of dispersion angles, while the axis of symmetry at $w\sim -2.5$ is that of the map of $N_\text{eff}$. These two axes differ because $\Delta\zeta \ne 0$.
    }
    \label{fig:holo4-003_fit}
\end{figure}

\begin{table}
\caption{Geometric characteristics of the 5 final holograms.}
    \centering
\begin{tabular}{lrrrr}
\toprule
       Ref \# &     $A'B'$ [mm] & $\Delta \zeta$ [mm] & $\alpha_0$ [deg] & $\chi^2_{\rm red}$ \\
\midrule
 Holo\#4-001 &  $19.4 \pm 0.2$ &   $-0.61 \pm 0.02$ &          $-0.38$ &              $4.7$ \\
 Holo\#4-002 &  $19.4 \pm 0.2$ &   $-0.73 \pm 0.03$ &          $-0.54$ &              $1.4$ \\
 {\bf Holo\#4-003} &  $19.4 \pm 0.2$ &   $-0.74 \pm 0.03$ &          $-0.56$ &              $1.5$ \\
 Holo\#4-004 &  $19.4 \pm 0.3$ &   $-0.64 \pm 0.04$ &          $-0.31$ &              $2.3$ \\
 Holo\#4-005 &  $19.4 \pm 0.2$ &   $-0.57 \pm 0.03$ &          $-0.30$ &              $1.6$ \\
\bottomrule
\end{tabular}
\tablefoot{$\Delta \zeta=\zeta_A - \zeta_B$ and $\alpha_0$ characterize the non-parallelism of A and B with respect to the w-axis in Fig. \ref{prod-holo}.
    }
    \label{tab:holo_geometry}
\end{table}

The mechanical equipment of the optical bench allowed us to scan the surface of the holograms coarsely and finely within the $(w,l)$ plane to study their dispersion properties.
For each position, we measured the dispersion angle $\alpha(w,l)$
and the separation between the zeroth-order centroid and the $546\,$nm $Hg$ line, from which the effective groove density $N_{\rm eff}(w,l)$ was derived using the known $D_{\rm CCD}$ value.
Using the hologram geometrical model from Paper 1, we fitted the
$(w,l,\zeta)$ positions of sources $A$ and $B$ (Fig. \ref{prod-holo}) to reproduce the maps $\alpha(w,l)$ and $N_{\rm eff}(w,l)$ from Fig. \ref{fig:holo4-003_fit}. 
To lift the degeneracy along $\zeta$, $\zeta_B$ was fixed to the CCD distance (202.3 mm), and only $\Delta \zeta = \zeta_A - \zeta_B$ was fitted.
The uncertainty on $\alpha(w,l)$ is estimated at $\sim 1\%$ while the
$N_{\rm eff}(w,l)$ uncertainty is dominated by the dispersion among the positions of the different $Hg$ lines.

\subsubsection{Geometric characteristics of the recorded holograms: true positions of the interfering sources}

\begin{figure}
    \centering
    \includegraphics[width=0.45\textwidth]{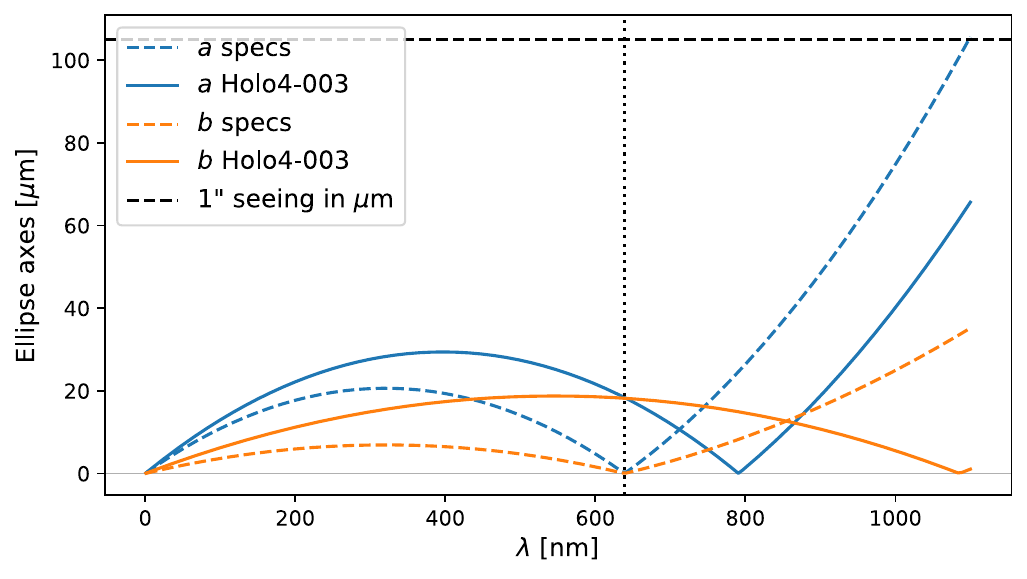}
    \caption{Major (a) and minor (b) axes sizes of the elliptical spot due to the defocusing of the spectrograph with respect to wavelength. The theoretical spot sizes expected from the maker's specifications are represented with dashed lines, while the Holo\#4-003 spot sizes, deduced from its measured parameters in Table \ref{tab:holo_geometry}, follow continuous lines. The dashed horizontal line near the top shows the size of a spot of 1 arcsec on the AuxTel detector.
    }
    \label{fig:holo4-003_focus}
\end{figure}

Results for Holo\#4-003 shown in Fig. \ref{fig:holo4-003_fit} are summarized in Table~\ref{tab:holo_geometry} along with the results for the other holograms.
For all five holograms, the geometric model fits the data well, as indicated by the reduced $\chi^2$ values.
The separation $A'B'$ is identical for all holograms, and the average dispersion angles $\alpha_0$, corresponding to the angle between $A'B'$ and the axis $w$, vary by less than $0.1^\circ$,
demonstrating good manufacturing repeatability. The fitted $\Delta \zeta$ values are consistent across holograms but differ from zero ($\sim 0.6mm$), indicating a slight non-orthogonality between the recording plate and the light beams. This explains the $\sim 5.5mm$ offset between the symmetry axes of the dispersion-angle and line-density maps in Fig. \ref{fig:holo4-003_fit}.
Using geometric optics, we computed the defocused spot sizes parallel and perpendicular to the dispersion for both an ideal hologram (designed with $D_{\mathrm{CCD}}=200\,$mm, $AB=20\,$mm and $\Delta \zeta = 0$) and Holo\#4-003 (Fig. \ref{fig:holo4-003_focus}). While perfect focus is expected at $\lambda=\lambda_R=639\,nm$ for the ideal case, Holo\#4-003 reaches optimal spectral resolution near $\lambda=800\,nm$.
In all cases, the focal spot provided by the optics is significantly smaller than the spot of a star scattered by AuxTel's seeing, indicating that the $\sim 0.6mm$ $\zeta$-offset does not degrade the hologram performance.

\subsection{Resolution of the AuxTel spectrograph}
The measured main characteristics of the final hologram from a geometric point of view are as described in paper 1.
Measurements performed on the test bench confirmed the good focus and that the separating power of the hologram is better than $1\,$nm at $\lambda=577\,$nm (Fig. \ref{fig:spectrogram_hg_ar}).
This result is consistent with the theoretical resolution expected from a standard ideal spectrograph with a periodic grating of 150 lines/mm crossed by a light beam with a “donut” cross-section measuring 11 mm in external diameter and 4 mm in internal diameter, corresponding to the projection of the AuxTel pupil onto the hologram plane.

In paper 1, we established the expression for the spectral resolution of a slitless holographic spectrograph (\cite{holospec1}, Eq. (15)), assuming that it is limited by seeing rather than by its maximum theoretical resolution.
This resolution is a function of $\sigma_0(\lambda)$,
the spatial dispersion of light on the focal plane expected for a monochromatic point source:
\begin{equation}
    R(\lambda,\sigma_0(\lambda)) = \frac{\lambda}{\Delta\lambda_\text{min}} = \frac{D_{\mathrm{CCD}}}{f \sigma_0(\lambda)} \frac{\lambda N_\mathrm{eff}}{[1-(\lambda N_\mathrm{eff})^2]^{3/2}}.
    \label{eq:resoltheor}
\end{equation}

where $\Delta\lambda_\text{min}$ is the separating power of the instrument (depending on $\lambda$) and $f$ is the focal length.
Thanks to good focus and the absence of optical distortion, and thanks to the high pixel sampling of AuxTel, $\sigma_0(\lambda)$ is dominated by atmospheric turbulence quantified by $\sigma_0=seeing/2.35$, and therefore
\begin{equation}
    R(\lambda,\sigma_0) \sim 272.\left[\frac{\sigma_0}{1\, {\rm arcsec}}\right]^{-1}\left[\frac{\lambda}{1\,\mu \text{m}}\right]
    \sim 639.\left[\frac{seeing}{1\, {\rm arcsec}}\right]^{-1}\left[\frac{\lambda}{1\,\mu \text{m}}\right].
    \label{eq:resolution}
\end{equation}
Assuming a seeing of $1\,$arcsec,
this resolution varies from 220 (at $\lambda=350\,$nm) to 640 (at $\lambda=1050\,$nm).
For the other diffraction orders ($p=2$ and $3$), the resolution is estimated with the same expressions, multiplied by the diffraction order $p$.

\subsection{Transmission properties}
\label{Sect:transmission}
\begin{figure} 
\begin{center}
\includegraphics[width=9.cm]{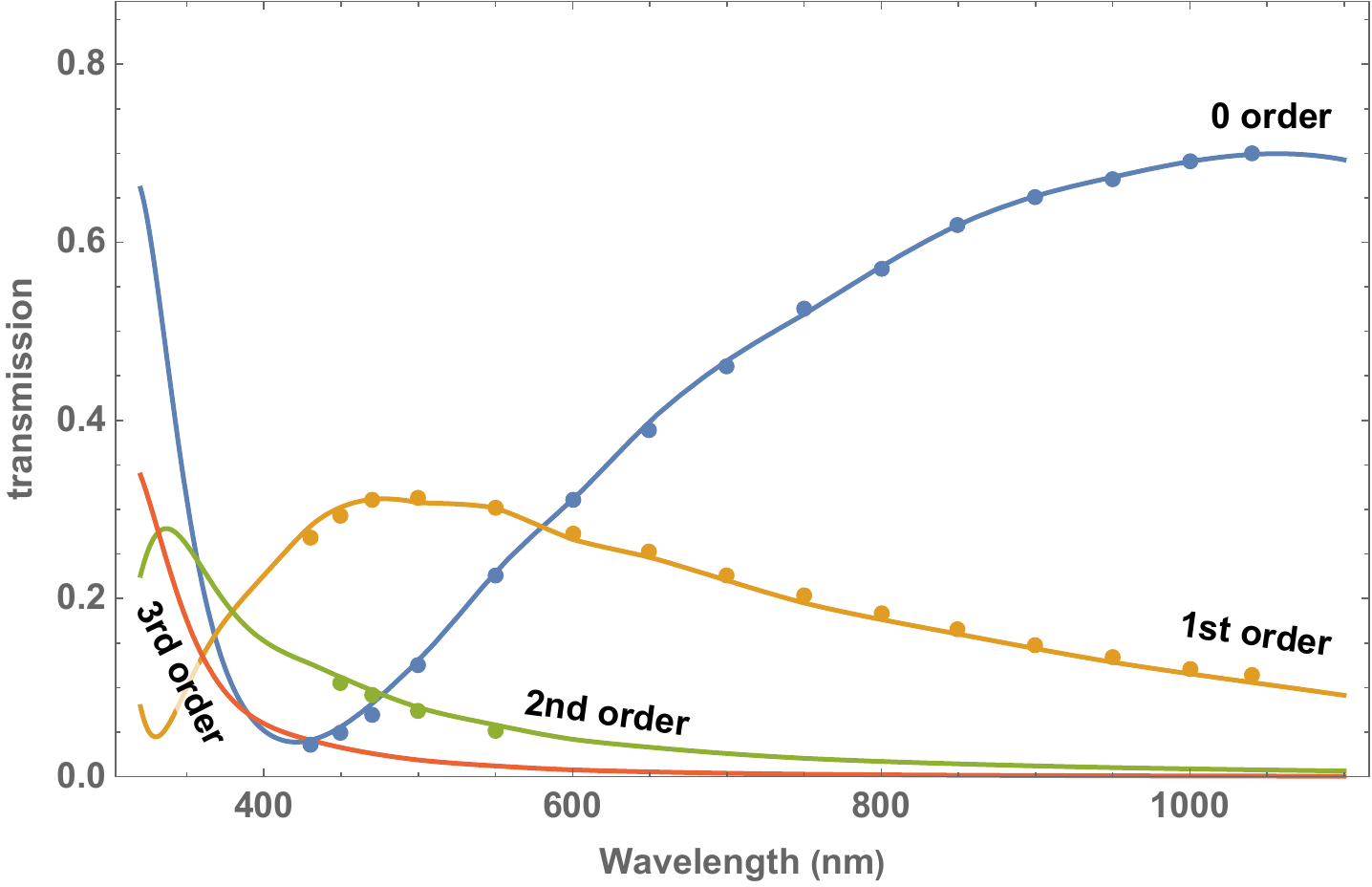}
\end{center}
\caption[] 
{
Transmissions $\eta_\text{p}(\lambda)$ of the different diffraction orders p of hologram Holo\#4-003.
The points represent the measurements from our test bench (error bars are smaller than the dots).
The solid lines indicate the theoretical transmissions $\eta_\text{p}(\lambda)$ for $p=0,1,2$ and $3$,
corresponding to an emulsion whose characteristics provide the best fit to the test bench measurements.
The transmissions take into account the residual absorption of the anti-reflective coated input surfaces.
}
\label{fig:simu-effic}
\end{figure}

Using the optical bench with the monochromator (Sect. \ref{Sect:bench}) we measured integrated fluxes in a narrow wavelength band without contamination
(see the points in Fig. \ref{fig:simu-effic}).
The measured ratios of the second order to the first order transmissions are plotted in red in Fig. \ref{ratios-orders}.
\label{Sect:bestfit}

We fitted the model described in Sect. \ref{Sect:model}, with the optical-bench transmission measurements (points in Fig. \ref{fig:simu-effic}) to determine the emulsion and exposure parameters of the final hologram. As noted in Sect. \ref{Sec:degeneracies}, several parameter combinations can reproduce the measured transmission curves. However, all acceptable solutions yield the same periodic phase-modulation pattern (Fig. \ref{sensitivity}, bottom), which governs the diffraction efficiencies at all wavelengths and orders. We therefore adopt the solution closest to the manufacturer’s specifications, corresponding to a beam intensity ratio of $\sim 1/3$ and a maximum index modulation of $\sim10\%$.
The curves in Fig. \ref{fig:simu-effic} show the model predictions using the parameters listed in Table \ref{tab:param} and the refractive index modulation plotted in Fig. \ref{modulation-vs-lambda}.
The agreement is good for all measured diffraction orders.

\begin{table}
\caption{Parameters of the final hologram process.
}
\begin{center}
\scalebox{0.95}{\begin{tabular}{cc}
\hline
manufacturer's specifications & \\
\hline
Emulsion thickness          & $5\,\mu$m \\ 
Recording wavelength        & $\lambda_\text{R}=639\,$nm \\
$E_0$                       & $600\,\mu\text{J/cm}^2$ \\
\hline
fitted parameters & \\
\hline
$\Delta_\text{max}(\lambda_\text{R})$   & $550\,$nm \\
$\gamma$                    & $3.4$ \\
max. index modulation at $\lambda_\text{R}$       & $0.1$ \\
Average deposited power $\overline{E_\text{rec}}$     & $342\,\mu\text{J/cm}^2$ \\
Beam intensity ratio               & $1/3$ \\
absorption of the emulsion  & $2\%$ \\
\hline
\end{tabular}}
\label{tab:param}
\end{center}
\end{table}

This model allows extrapolation toward shorter wavelengths and estimation of the contribution of the third diffraction order, which is not accessible on the test bench but appears on the AuxTel detector beyond the infrared end of first-order diffraction.
Further refinement of this empirical model is beyond the scope of this work. Nevertheless, the level of agreement is satisfactory given the simplifying assumption of a locally periodic grating illuminated by a parallel beam, which is only valid to first order.

\section{On-sky validation and commissioning}
\label{Sect:discussion}

\subsection{Tests at the Pic-du-Midi observatory}
\label{Sect:pic}
During the optimization process, we performed measurements on sky to verify the optical properties of one of our prototypes in real conditions.
From 14 to 16 February 2019, one of our first sealed holograms have been mounted on the 1 meter f/17.5 telescope of the Pic du Midi, a telescope whose geometry is very close to the geometry of the LSST auxiliary telescope (1.2m, f/18).
This versatile instrument
allowed us to put the holographic element at the same distance $D_\text{CCD}$ to the focal plane than it should be in AuxTel. Therefore, the geometrical configuration of the converging beam was almost identical to the one expected with AuxTel, providing us the capacity to perform a completely realistic test on sky.
The images were focused on a 2048x2048 CCD, with pixel size of $24\,\mu$m (corresponding to a pixel scale of $0.23$ arcsec).
Thanks to this instrument, we could complete the test-bench operations with two complementary tests:
\begin{itemize}
    \item 
    Testing the effect of tilting the hologram by $1\degree$ with respect to the optical axis; this is a request from AuxTel's spectrograph specifications to avoid the superimposition of ghosts near the spectrum.
    \item
    More extensive exploration of the fiducial volume of the hologram, {\it i.e.} the domain of the hologram positions relative to the beam where the optical functions are near the optimum within a few percent.
\end{itemize}
The result is that there is no measurable impact when tilting the hologram; furthermore examination of the spectrum profile widths from raw images confirmed us that the first order stellar spectrum profile was stable whatever be the distance between the CCD and the hologram within the interval $120 < D_{CCD} < 200\,$mm.

\begin{figure}
\begin{center}
\includegraphics[width=0.9\linewidth]{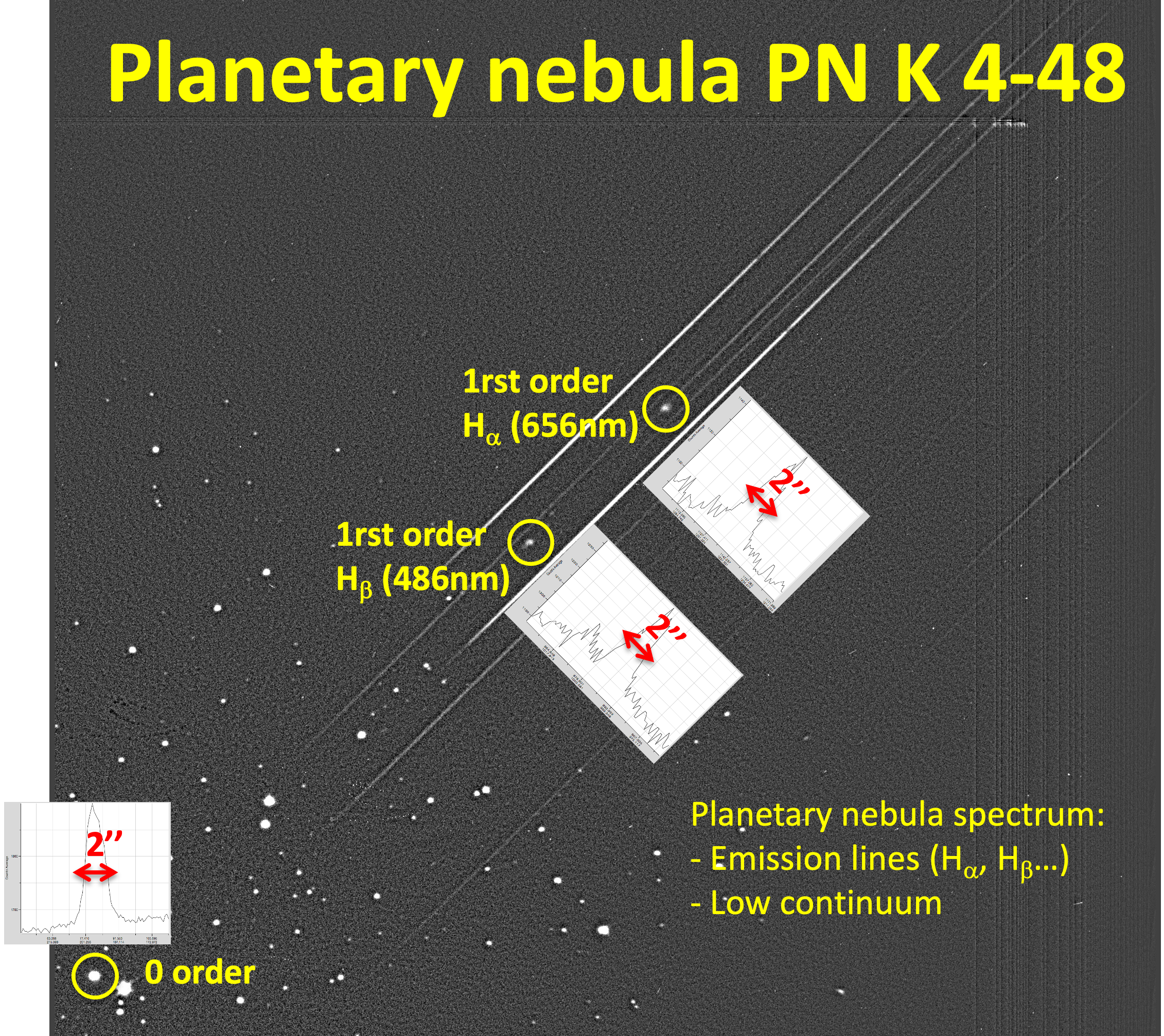}
\caption{Spectrogram of the planetary nebula PN K 4-48 obtained with the 1-meter telescope at the Pic du Midi, showing the quasi-monochromatic $H_{\alpha}$ and $H_{\beta}$ emission lines. The insets show the spectral line cross-section at zero order and first diffraction order for $H_{\alpha}$ and $H_{\beta}$. The widths are the same for all three.}
  \label{fig:picmidi}
\end{center}
\end{figure}

Figure \ref{fig:picmidi} shows the spectrogram of a planetary nebula that confirms the quality of the focus at all wavelengths,
showing resolution only limited by the seeing.

\subsection{Commissioning of the AuxTel spectrograph and improvements}
Hologram holo$\#4$-003 has been installed on the disperser wheel of the auxiliary telescope since February 15, 2021,
and data has been taken since then.
The purpose of commissioning of the hologram after installation on the spectrograph was to check the performance of the system --- focusing and transmission ratios --- and to improve it.
Below we present confirmatory measurements, along with additional checks not accessible on the laboratory test bench.
The first results on the atmosphere will be published shortly in a separate article.

\subsubsection{Transmission ratios measured on the sky with AuxTel}
\label{subsec:transauxtel}
\begin{figure}
    \centering
    \includegraphics[width=0.45\textwidth]{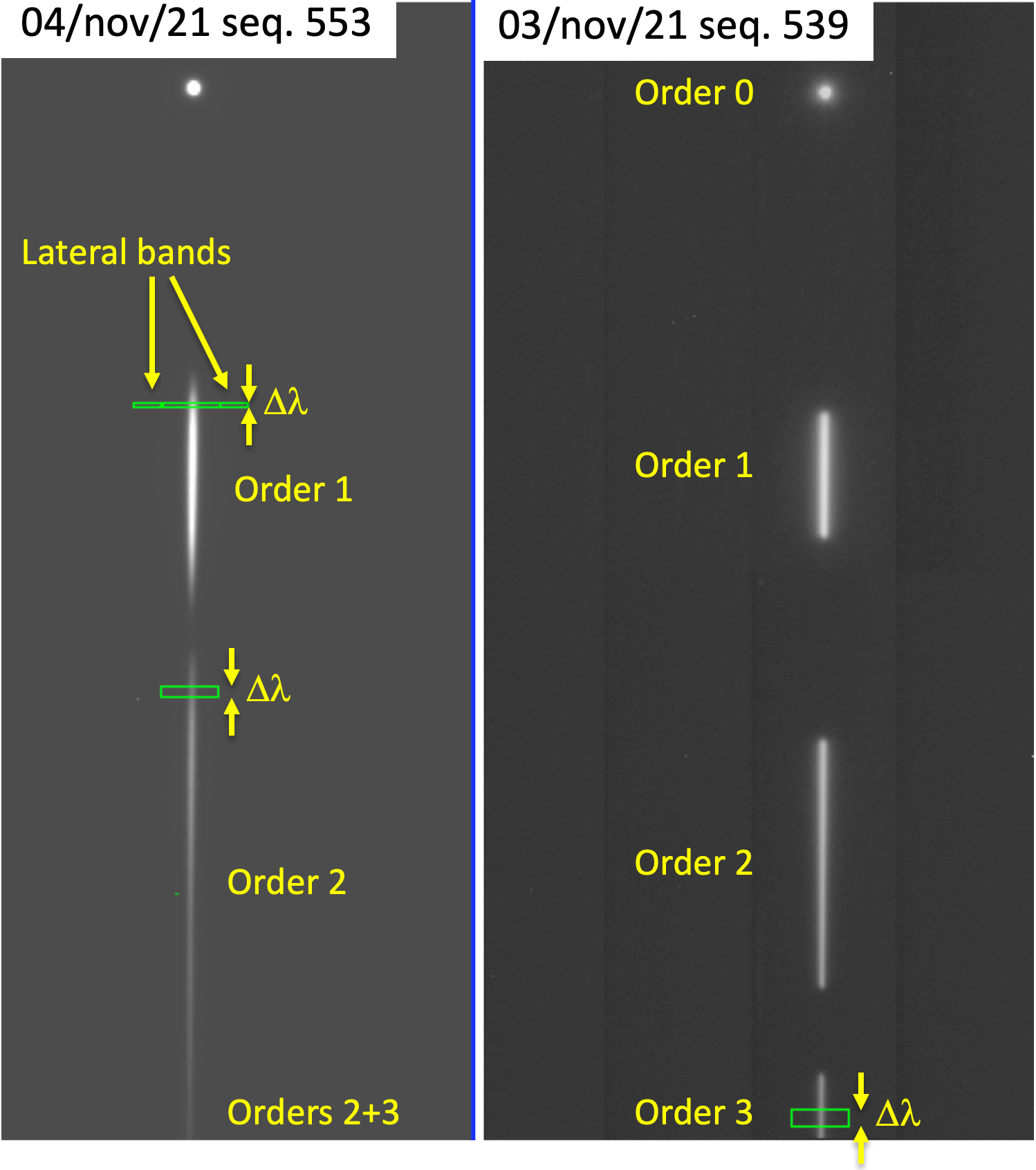}
    \includegraphics[width=0.45\textwidth]{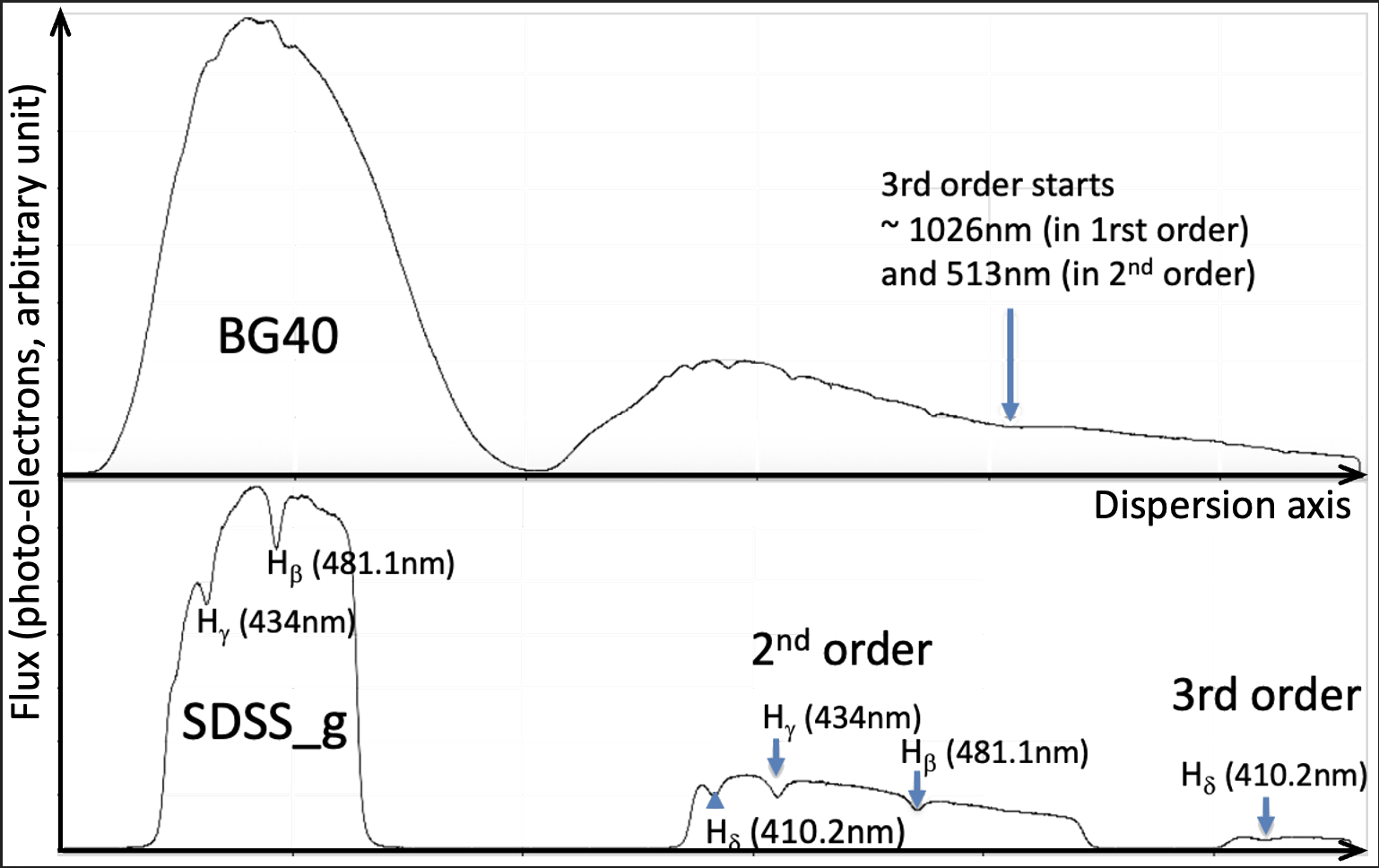}
    \caption{Principle of the measurements of the diffraction order ratios.
    Top: On the left, the spectrogram (spread from top to bottom) is obtained after passing through the BG40 filter, while on the right the spectrogram is obtained after passing through the narrower SDSSg filter, which allows us to isolate the 3rd-order contribution.
    The green rectangles show the integration domains of the fluxes.
    Bottom: the projections from the spectrograms above, showing some of the star's spectral characteristics, and the limits of the filters' bandwidths.
    }
    \label{fig:ratiosky}
\end{figure}

Rather than absolute transmissions, the ratios, which are easier to measure accurately, are the most useful for Spectractor software settings.
But first, we determined the wavelength range of our spectrograph.
Using stars of different colors, we were able to establish the limits of the detectable spectrum as $[345,1050]\,$nm.
To update and complement the optical bench measurements in the UV region, we analyzed the different orders of a continuous spectrum, that of $\mu$Col (HD38666 - $(\alpha,\delta)_{J2000}=(5h46m00s, -32\degree 18'23.2'')$), a hot UV-bright star with magnitude $V=5.2$.
To obtain transmission ratios, we used red-blocking filters (BG40 and SDSSg) to avoid mixing orders.
The images were taken with an airmass close to one, thus avoiding any chromatic differential refraction from the atmosphere.
The relative flux density is then integrated within a central sliding window of width $\Delta \lambda=7.3\,$nm (see Fig. \ref{fig:ratiosky}-up) along the dispersion axis -- equivalent to 20 pixels for the first dispersion order, $\sim 40$ pixels for the second order and $\sim 60$ pixels for 3rd order-- and 200 pixels in the transverse direction. Two lateral sliding windows are used to estimate the background, with the same $\Delta \lambda$ and 100 pixels in the transverse direction. The sum of both lateral bands is subtracted to the central band signal to correct from sky background.
The order ratios are then estimated by the density ratios at the same wavelengths.
We checked the stability of this estimate by varying the width $\Delta\lambda$ of the intervals.

We took into account a particular effect caused by the dispersion of light at each wavelength due to seeing;
the flux integrated in a given bin is therefore contaminated by adjacent wavelengths, resulting in an asymmetric bias when the flux gradient is steep \citep{eddington_1913}, which can occur with stellar spectra, particularly in the UV. 
This effect differs between orders 1 and 2 due to their different dispersions and transmission slopes.
We estimate this systematic effect by simulating a displacement of the source of $0.8\,$arcsec on the red side of the spectrum;
the resulting impact on the transmission ratio is shown by the black dashed curve in Fig. \ref{ratios-orders}. An additional uncertainty arises from the determination of the zeroth-order position; its effect, assuming a $\pm 1$ pixel uncertainty, is shown by the blue error bars. Other sources of uncertainty (wavelength calibration, emulsion scattering, CCD gain variations, spectrogram tilt, and Poisson noise) are negligible.

The same technique was used to estimate the ratio of 3rd- and 2nd-order transmissions, based on the spectrogram taken through the SDSSg filter, which is sufficiently narrow to enable their separation; but this experimental determination is only possible in the $[400,420]\,$nm range, limited by the SDSSg blue cut-off (at $400\,$nm) and the CCD edge (reached for 3rd order at $\lambda\sim 420\,$nm), beyond which light is no longer detected.

\subsubsection{Synthesis of all transmission measures}

\begin{figure} 
\begin{center}
\includegraphics[width=8.cm]{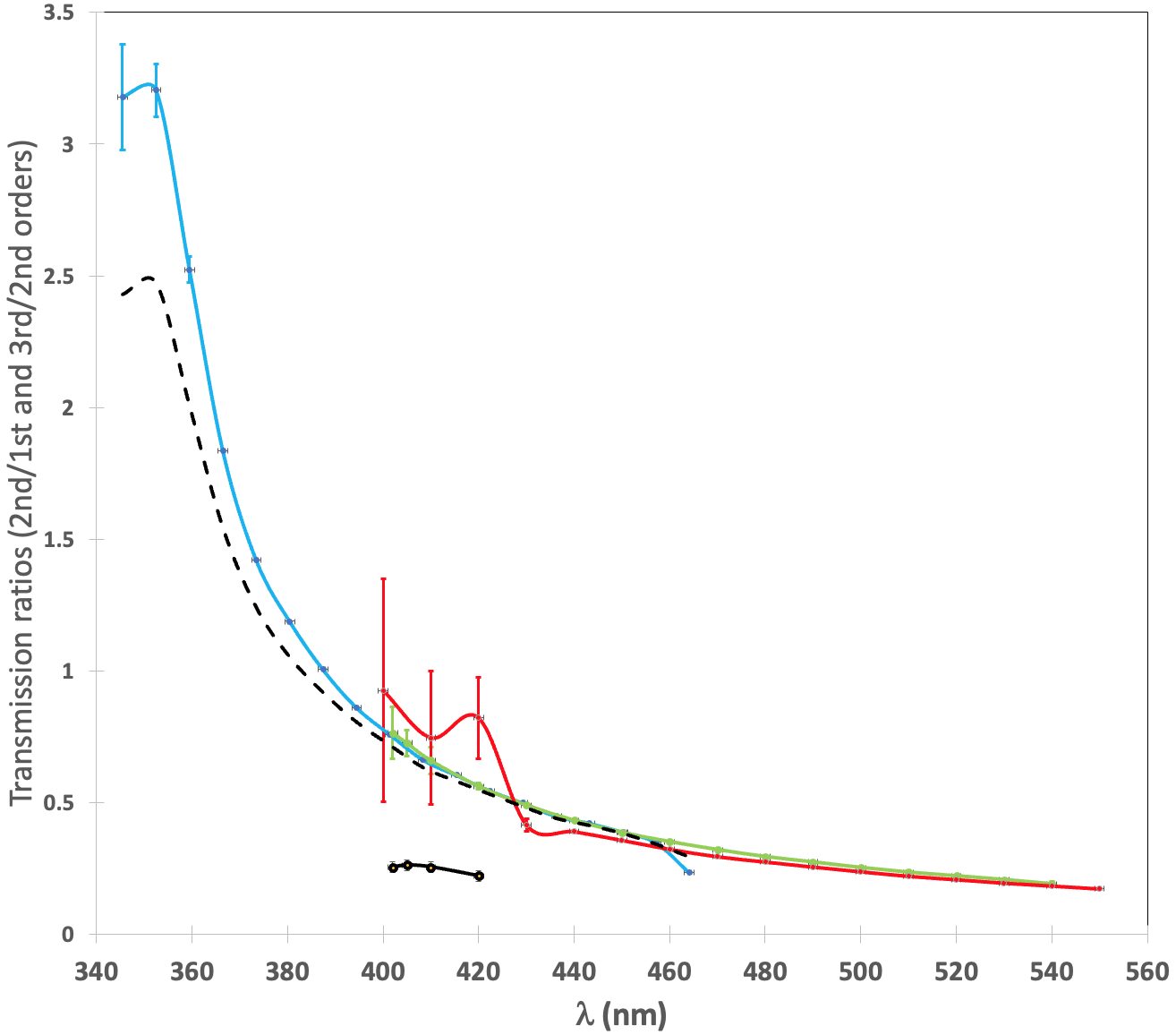}
\end{center}
\caption[] 
{
Measured 2nd/1st order transmission ratios of Holo\#4-003 with the test-bench (red), with a spectrum obtained through BG40 filter (blue), and with a spectrum obtained through the SDSSg filter (green).
The blue error bars are estimated from the uncertainty of the zero order position.
The black dashed line indicates the values obtained assuming a source shift of $0.8\,$arcsec toward the red part of the spectrum
(see text). 
The short black line (between $400$ and $420\,$nm) shows the measured 3rd/2nd order transmission ratio.
}
\label{ratios-orders}
\end{figure}

Figure \ref{ratios-orders} combines hologram transmission ratios measured on the optical bench and on sky, showing excellent agreement above $\lambda=460\,$nm.
BG40-filter data extend the coverage where bench measurements are flux-limited.
In addition to these techniques, on-sky airmass-regression methods implemented in Spectractor, combined with the known telescope throughput, provide an independent and consistent determination of the hologram transmission.
A CBP-based (Collimated Beam Projector), tunable monochromatic illumination system \citep{CBP_2025} will soon enable regular, bench-equivalent transmission measurements under operational conditions; these calibration techniques and their applications will be presented in forthcoming papers.

\subsubsection{Field star masking}
\begin{figure} 
\begin{center}
\includegraphics[width=1.5cm]{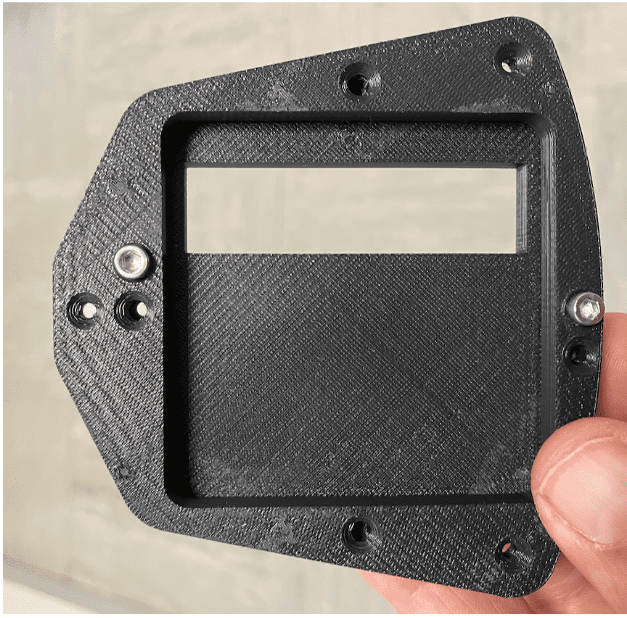}
\includegraphics[width=7.cm]{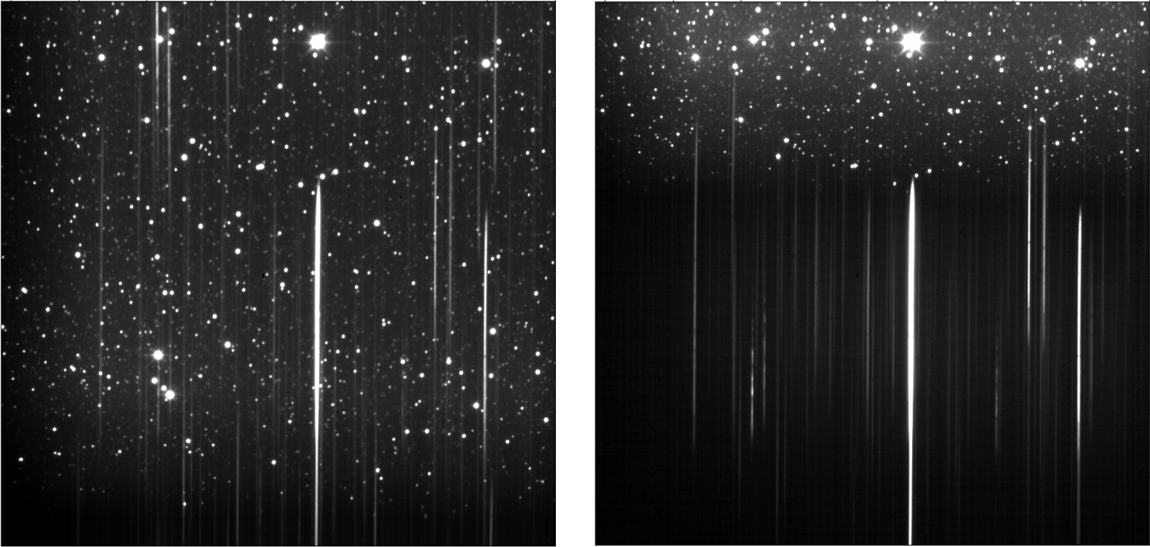}
\end{center}
\caption[] 
{
Left: the mask inserted on the entrance side of the hologram.
Middle: image obtained with the unmasked hologram: the direct image of the stars and sky background are superimposed on the spectrum of interest. On the right, the mask has been inserted to eliminate the direct image. All that remains visible in the fully masked area is diffracted light: stellar spectra and, at each point, integrated sky background contributions at all (non-zero) orders and wavelengths.
}
\label{fig:mask}
\end{figure}

We integrated a rectangular mask into the hologram frame, designed to mask the direct image of background stars. Figure \ref{fig:mask} shows the effectiveness of this device, which not only eliminates the contamination of field stars, but also significantly reduces the sky background over the entire spectrum.
The width of this mask ($15mm$) is such that the beam, whose cross-section is approximately $11mm$ in diameter at the hologram, is not intercepted. On the other hand, the mask is not limited in the direction perpendicular to the dispersion, so as to enable good estimation of the sky background on either side of the band containing the spectrum of the source under study.

Because the mask is far from the focal plane, the sky background at a given position on the detector is the integral of the sky spectrum over a wide range of wavelengths, resulting in a smoothly varying background that is easy to subtract during analysis.

\section{Conclusion and perspectives}
\label{Sect:conclusion}

We have described the optimization of the production process of a thin holographic optical element designed as a single planar dispersive element.
We have also succeeded in accurately modeling the production process and the characteristics of our hologram, which is essential for the accuracy of standard star spectrophotometry.
This hologram is inserted in the path of a converging beam, like an ordinary filter, without additional optics, providing good focusing from $345\,$nm and $1050\,$nm on an on-axis detector.
After five years of operation on AuxTel, this element has demonstrated a wavelength resolution superior to that of a periodic grating and a high diffracted light throughput (up to $32\%$, close to the theoretical maximum possible).
Coupled with Spectractor and with minimal precautions to avoid contamination by field stars, it enables the high performance slitless spectrophotometry needed to study atmospheric transmission and is now in use as part of the Rubin-LSST survey.

The AuxTel configuration is common to many telescopes, so this technique could be generalized to transform any imaging telescope into a high-performance, low-cost slitless spectrograph. However, targets must remain near the optical center of the holographic element for an optimal use.

\begin{acknowledgements}
The authors would like to thank Kirk Gilmore, Robert Lupton, and Christopher Stubbs for their contributions to the discussions that helped refine the specifications for the holographic solution after the initial trials.
We would like to thank Patrick Ingraham, Erik Dennihy, Elana Urbach, and the team of observers for their contributions to the commissioning of the hologram, as well as Roberto Tighe and his team of opticians at the Vera Rubin Observatory for installing the optical elements on the AuxTel telescope.
Our thanks also go to Olivier Perdereau for his participation in the process of optimizing exposure times and emulsion type, to Merlin Fisher-Levine for his involvement in the intricacies of computer processing, and to Benoît Sassolas for suggesting the involvement of the Advanced Materials Laboratory (LMA) in Lyons to apply a broadband multi-layer coating to the faces of the hologram.
\end{acknowledgements}

\bibliographystyle{aa}
\bibliography{citations-holospec2}

\appendix
\section{The optical function of the hologram}
\label{appendix_math}
\subsection{The recording}
Figure \ref{prod-holo} shows the notation convention for the holographic recording.
We will keep the same frame for the image restitution (changing $D_R$ with $D_{CCD}$).
Let $\lambda_R$ be the recording laser wavelength, $D_{R}$ the distance between the source plane and the hologram plane, and $d_R$ the distance between the A and B sources;
the spatial term of the complex illumination from the reference point-source $A(-d_R/2,0,D_R)$ (similarly for the object point-source $B(d_R/2,0,D_R)$) at a position $M(w,l,0)$ in the hologram plane is given by
\begin{equation}
\mathbf{U}_A(w,l,\lambda_R) =U_A(\lambda_R) \frac{e^{ik_R.r_{AM}}}{r_{AM}},
\label{spherical-wave}
\end{equation}
where $k_R=2\pi /\lambda_R$,
$r_{AM}=\sqrt{D_{R}^2+(w+d_R/2)^2+l^2}$ is the distance from $A$ to $M$, and $U_A(\lambda_R)$ is the source amplitude. In the holographic plane,
the sensitive emulsion records the modulated intensity pattern resulting from the interference of the two point-sources $A$ (reference) and $B$ (object), given by
\begin{equation}
I_{rec}(w,l)= \left| \mathbf{U}_A(w,l,\lambda_R) + \mathbf{U}_B(w,l,\lambda_R) \right|^2.
\label{interference}
\end{equation}
In section \ref{Sect:bestfit}, we used the ratio of beam intensities $R=U_B^2/U_A^2$ as a parameter. It is related to the interference contrast $C$ by the relation $C=2\sqrt{R}/(1+R)$.
After exposure during $t_{exp}$, followed by the chemical process,
the emulsion response to the exposure spatial modulation $E_{rec}(w,l)=I_{rec}(w,l)\times t_{exp}$ is converted into an 
optical density modulation in the case of amplitude holograms or into an optical index modulation after bleaching processing in the case of  phase holograms.

In our case, since the spatial modulation is a pure refraction index modulation,
the local complex 2D transmission function of the emulsion
$\mathbf{H}(w,l,\lambda)$ for a light-beam of wavelength $\lambda$
is derived from the local optical path supplement map $\Delta(E_{rec}(w,l),\lambda)$:
\begin{equation}
\mathbf{H}(w,l,\lambda)=e^{2i\pi \Delta(E_{rec}(w,l),\lambda)/\lambda}.
\end{equation}
Here $\Delta(E_{rec}(w,l),\lambda)$ depends on the local energy $E_{rec}(w,l)$ deposited during the recording and also on the wavelength as soon as the refraction index modulation varies with $\lambda$.

\subsection{Image restitution}
At the entrance to the hologram, located at $D_{CCD}$ (which differs from $D_R$) upstream of the detector plane, the amplitude of a spherical wave with wavelength $\lambda$ converging at point
$S(w_S,0,D_{CCD})$
at position $M(w,l,0)$ on the hologram is
\begin{equation}
\mathbf{U}_{S}(w,l,\lambda) =U_{S}(\lambda) \frac{e^{-ik.r_{MS}}}{r_{MS}},
\end{equation}
where $k=2\pi /\lambda$,
$r_{MS}=\sqrt{D_{CCD}^2+(w-w_S)^2+l^2}$ is the distance from $S$ to $M$,
and $U_{S}(\lambda)$ is the image amplitude,
similarly to Eq. (\ref{spherical-wave});
the amplitude after crossing the hologram is given by
\begin{equation}
\mathbf{U}_{S}(w,l,\lambda). \mathbf{H}(w,l,\lambda),
\end{equation}
where the transfer function $\mathbf{H}$ includes the telescope's pupil function.
In the case of AuxTel, it corresponds to a circular aperture with a diameter of $11mm$, centered on $A'$.
Since the incident wave is converging,
the amplitude of the image at position $(w,l,D_{CCD})$ (in the focal plane) simply results from the Fraunhofer diffraction

\begin{eqnarray}
\lefteqn{\mathbf{U}_{CCD}(w,l,\lambda)\approx} \\
& & \frac{U_{S}(\lambda).e^{\frac{i\pi(w^2+l^2-w_S^2)}{\lambda D_{CCD}}}}{i\lambda D_{CCD}^2}
\int\!\!\int_{-\infty}^{+\infty}\mathbf{H}(u,v,\lambda)
e^{\frac{-2i\pi((w-w_S).u+l.v)}{\lambda D_{CCD}}}\ud u\ud v,  \nonumber
\end{eqnarray}
which is conveniently written as a Fourier transform, easy to estimate through the FFT algorithms.
The image intensity is then given by the expression
\begin{equation}
\mathbf{I}_{CCD}(w,l,\lambda)=\frac{U_{S}(\lambda)^2}{\lambda^2 D_{CCD}^4}\left| FT[\mathbf{H}]_{\left( \frac{w-w_S}{\lambda D_{CCD}},\frac{l}{\lambda D_{CCD}}\right)}\right| ^2.
\end{equation}
Our notation means that the Fourier transform has to be estimated at spatial frequencies
$\left( \frac{w-w_S}{\lambda D_{CCD}},\frac{l}{\lambda D_{CCD}}\right)$ where the coordinate system $(w,l)$, attached to the focal plane, has its origin at the intersection of the $\zeta$-axis with that same plane.
Note that the intensity pattern is only a function of the relative $(w,l)$ position with respect to the position of the zero-order image $S$ in the CCD plane. This remains true as long as the paraxial approximation is valid.
In the particular case of AuxTel, the pupil function to be integrated into the transfer function $\mathbf{H}$ corresponds to a circular aperture with a diameter of $11mm$, centered on $A'$.

\end{document}